\documentclass[aps,pre,twocolumn,amsmath,amssymb,superscriptaddress,showkeys,floatfix]{revtex4}

\usepackage{graphicx}
\usepackage{graphics}
\usepackage{bm}
\usepackage[usenames]{color}
\usepackage{colordvi}
\usepackage{units}
\usepackage{bbm}
\usepackage{enumitem}
\usepackage{calc}
\usepackage{verbatim}
\usepackage{multirow}
\usepackage[utf8]{inputenc}
\usepackage{amsmath}
\usepackage{amssymb}
\usepackage{latexsym}
\usepackage{mathtools}
\usepackage{epsfig}
\usepackage{epstopdf}
\usepackage{rotating}
\usepackage[english]{babel}

\usepackage{algorithm}

\usepackage{algpseudocode}

\usepackage{amsfonts}
\usepackage{placeins}
\usepackage{braket}
\usepackage{natbib}
\usepackage{tabularx}

\newcommand{\one}{\Ket{O}}
\newcommand{\two}{\Ket{OO}}
\newcommand{\three}{\Ket{OOO}}

\newcommand{\oneh}{\Ket{O^H}}

\newcommand{\twoh}{\Ket{OO^H}}
\newcommand{\twohh}{\Ket{O^HO^H}}

\newcommand{\threeh}{\Ket{OOO^H}}
\newcommand{\threehh}{\Ket{OO^HO^H}}
\newcommand{\threehhh}{\Ket{O^HO^HO^H}}

\newcommand{\icrac}{I_{CRAC}}
\newcommand{\icracstat}{I^{stat}_{CRAC}}
\newcommand{\icrach}{I_{CRAC}^{mix}}

\newcommand{\no}{N_{Z_1}}
\newcommand{\noo}{N_{Z_2}}
\newcommand{\nooo}{N_{Z_3}}

\newcommand{\noh}{N_{\Ket{O^H}}}

\newcommand{\nooh}{N_{\Ket{OO^H}}}
\newcommand{\nohoh}{N_{\Ket{O^HO^H}}}

\newcommand{\noooh}{N_{\Ket{OOO^H}}}
\newcommand{\noohoh}{N_{\Ket{OO^HO^H}}}
\newcommand{\nohohoh}{N_{\Ket{O^HO^HO^H}}}

\newcommand{\hzwei}{H_2O_2}

\newcommand{\cazweiplus}{Ca^{2+}}
\newcommand{\oh}{O^H}
\newcommand{\otot}{O^{tot}}

\newcommand{\stot}{S^{tot}}

\begin{document}

\title{Reaction-diffusion model for STIM-ORAI interaction: the role of ROS and mutations}

\author{Barbara Schmidt}
\email{schmidt@lusi.uni-sb.de}
\affiliation{Department of Theoretical Physics, Saarland University, 66041 Saarbrücken, Germany}
\affiliation{Department of Molecular Biophysics, Saarland University, 66421 Homburg, Germany}
\affiliation{Department of Biophysics, Saarland University, 66421 Homburg, Germany}
\author{Dalia Alansary}
\email{dalia.alansary@uks.eu}
\affiliation{Department of Molecular Biophysics, Saarland University, 66421 Homburg, Germany}
\author{Ivan Bogeski}
\email{ivan.bogeski@med.uni-goettingen.de}
\affiliation{Department of Biophysics, Saarland University, 66421 Homburg, Germany}
\affiliation{Molecular Physiology, Institute of Cardiovascular Physiology, University Medical Center Georg-August-University, 37073 Göttingen, Germany}
\author{Barbara A. Niemeyer}
\email{barbara.niemeyer@uks.eu}
\affiliation{Department of Molecular Biophysics, Saarland University, 66421 Homburg, Germany}
\author{Heiko Rieger}
\email{h.rieger@mx.uni-saarland.de}
\thanks{Corresponding author}
\affiliation{Department of Theoretical Physics, Saarland University, 66041 Saarbrücken, Germany}
\date{\today}

\begin{abstract}
Release of $\cazweiplus$ from endoplasmatic retriculum (ER) $\cazweiplus$ stores causes stromal interaction molecules (STIM) in the ER membrane and ORAI proteins in the plasma membrane (PM) to interact and form the $\cazweiplus$ release activated $\cazweiplus$ (CRAC) channels, which represent a major $\cazweiplus$ entry route in non-excitable cells and thus control various cell functions. 
It is experimentally possible to mutate ORAI1 proteins and therefore modify, especially block, the $\cazweiplus$ influx into the cell. On the basis of the model of Hoover and Lewis (2011) \cite{hoover2011}, we formulate a reaction-diffusion model to quantify the STIM1-ORAI1 interaction during CRAC channel formation and analyze different ORAI1 channel stoichiometries and different ratios of STIM1 and ORAI1 in comparison with experimental data. We incorporate the inhibition of ORAI1 channels by ROS into our model and calculate its contribution to the CRAC channel amplitude. We observe a large decrease of the CRAC channel amplitude evoked by mutations of ORAI1 proteins. 
\end{abstract}



\maketitle

\section{Introduction}
Temporally and locally controlled changes of the intracellular $\cazweiplus$ concentration drive a plethora of cellular functions \cite{berridge2000, carafoli2002, clapham2007}.
While electrically excitable cells utilize voltage gated $\cazweiplus$ channels to achieve rapid changes in intracellular $\cazweiplus$ concentration \cite{clapham2007}, immune cells require slower but long-lasting changes in intracellular $\cazweiplus$ concentration for activation and cytokine production \cite{falke2004, lewis2001}. In these cells activation of T-cell receptors and other PLC coulped receptors results in a depletion of intracellular $\cazweiplus$ stores (endoplasmatic reticulum, ER). 
This results in an increased intracellular $\cazweiplus$ concentration, which by itself is not sufficient to trigger long-term immune cell activation and translocation of the nuclear factor of activated T-cells (NFAT).
Thus extracellular $\cazweiplus$ needs to be ingested into the cell.

For long-lasting $\cazweiplus$ influx the information about the filling state of the ER has to be conveyed to ion channels in the plasma membrane, which then provide an entry pathway for extracellular $\cazweiplus$ \cite{parekh2005, hoth1992, derler2016, clapham2007, muik2008, luik2008}. The drop in ER luminal $\cazweiplus$ is sensed by stromal interaction molecules (STIM), which undergo a conformational change, multimerize and move to regions near the plasma membrane, so called plasma membrane junctions (PMJ \cite{wu2011, liou2007}). 
Here STIM1 proteins trap ORAI1 ion channel proteins diffusing within the plasma membrane \cite{liou2007, park2009, yuan2009, luik2008}.
Depending on the STIM1-ORAI1 stochiometry different ORAI1 conductance states are reached open and selectively conduct $\cazweiplus$ ions into the cell.

Although the electrophysiological correlate has been known as CRAC ($\cazweiplus$ release activated $\cazweiplus$) current since 1991, the exact protein composition of the complex has been a matter of debate. Several reports pointed towards dimeric ORAI1 channels at rest while tetrameric and hexameric stoichiometries have been proposed to form the ion conduction pore. Indeed recent results obtained from using concatenated constructs as well as the crystal structure of purified \textbf{\textit{Drosophila}} Orai1 point towards hexameric Orai1 channels as the predominant species underlying $\icrac$ \cite{li2016, yen2016, cai2016}.

While the amplitude of $\icrac$ is thus determined by the relative ratios of the STIM1 to ORAI1 protein levels \cite{hoover2011,derler2016,scrimgeour2009}, it is also modified by external factors. When immune cells enter an area of inflammation, they encounter environments rich in reactive oxygen species (ROS). Exposure to ROS prevents ORAI1 from being activated whereas preassembled STIM1-ORAI1 complexes are insensitive towards inhibition by ROS \cite{bogeski2010}.

In \cite{hoover2011} a Monod-Wynman-Changeux model is used to analyze the dependence on $\icrac$ on ORAI1 expression levels. In this model ORAI1 exists in two conformational states (open and closed) with four binding sites of STIM1 each, which results in ten channel states in total. Analysis of ORAI1 complexes with less binding sites disagreed with the experimental data. More binding sites have not been tested. 
With the assumption that STIM1 binds to ORAI1 with negative cooperativity the model provides a reasonable fit to $\icrac$. While for an ORAI1 expression level less than two (a. u.) the dominating state is the open ORAI1 conformation with four STIM1 bound $OS_4$, for higher ORAI1 expression levels the open states with three and two STIM1 bound dominate and the state $OS_4$ is nearly not occupied anymore \cite{hoover2011}. 

The stochastic reaction diffusion model of \cite{peglow2013} assumes that the ORAI1 proteins are already accumulated into tetramers and that these ORAI1 complexes can trap one to four STIM1 dimers. This leads to four different CRAC channel states with different current capacities contributing to the calculation of $\icrac$ and to CRAC channel currents reaching steady state values not earlier than about 300 s.  Furthermore the numbers of different CRAC channel states in dependence on the cooperativity factors $\alpha$ and $\beta$ is analyzed. It is found that for all tested configurations $(\alpha,\beta)$ the best ratio of STIM1 monomers to ORAI1 tetramers is 7.5. In addition the number of active channels depends not only on the total number of ORAI1 proteins but also on the chosen cooperativity. For negative cooperativity and a small number of ORAI1 proteins the tetrameric states are most dominant, for a higher number of ORAI1 proteins the single STIM1 bound states dominate.

In \cite{wu2014} the diffusional behavior of STIM1 and ORAI1 in PMJ regions is addressed. 
Single-particle tracking and photoactivation experiments are combined with Monte Carlo simulations to analyze STIM1 and ORAI1 diffusion at PMJ regions in resting cells as well as in activated cells. It is found that in resting cells STIM1 proteins follow Brownian motion and ORAI1 motility is subdiffusive. After ER-store depletion ORAI1 and STIM1 movement is mostly restricted to PMJ spots and the data show that both proteins move together as a complex. Furthermore, in activated cells more proteins are immobile than in resting cells.

The junctional regions between ER and plasma membrane are not static \cite{shen2011} but the ER is remodeled by STIM1 proteins. Activated STIM1 proteins are able to form elongated ER cisternae close to the plasma membrane, which increase in number and length during ER depletion \cite{sauc2015, wu2006}. The PM junctions seem to be pre-determined since the puncta structures appear repeatedly at the same spots \cite{malli2008}.

In \cite{melunis2016} an integrated particle system model is combined with stochastic modeling using a spatially heterogenous Gillespie algorithm to track single particles during CRAC channel formation.
The presented model allows to observe the dynamics of single molecules as in single molecule tracking experiments and allows to change rates of action according to the history of the molecules.

Using a similar setup as \cite{peglow2013} the diffusion of STIM1 complexes and the four different channels states is analyzed. Therefore the size of the PMJ is enlarged and the channel states do not rest at one single PMJ spot. In the beginning none of the species show sub-diffusive behavior but after 30 s the CRAC channel states move sub-diffusively with decreased diffusion rate, which is a result of restricting these complexes to the small PMJ spot. Together with this restriction the formation of STIM1-ORAI1 complexes results in differences of STIM1 movement before and after releasing $\cazweiplus$ from ER. 

While previous studies have used a reaction-diffusion system to model $\icrac$ \cite{hoover2011, kilch2013, peglow2013, melunis2016}, these models were based on a tetrameric CRAC channel configuration and a single junctional interaction region.

Recent studies analyzing the CRAC channel stoichiometries conclude that activated ORAI1 channels are present as hexameres. Yen et al. expressed hexameric concatemers of human ORAI1 \cite{yen2016}. Their measured currents reproduce the characteristics of CRAC channel current which gives evidences that CRAC channels follow a hexameric stoichometry. 
Single-molecule photobleaching experiments lead to the conclusion that ORAI1 forms dimers at resting state and that upon activation a mixture of dimers, tetramers and hexamers contributes to the total CRAC current \cite{li2016}.
The ORAI1 concatemer analysis of \cite{cai2016} finds that concatemers of different sizes (dimer to hexamer) all lead to significant  $\cazweiplus$ influx. Substitution of non-conducting  subunits into different places in the concatemers shows, that the conducting capability of the whole hexameric concatemer depends on the position of the substituted subunit. This leads to two differnt possibilities for the CRAC channels: a pure hexameric arrangement and an assembly as \glqq trimer-of-dimers\grqq.

The ROS-mediated CRAC channel inhibition is analyzed in \cite{alansary2016}. Next to FRAP measurements to find differences in the diffusion parameter of inhibited and non-inhibited ORAI1, FRET measurements analyze the interaction strength between STIM1 and  ORAI1 and between ORAI1 and ORAI1 for both WT and ROS-preincubated ORAI1. Furthermore, mutated ORAI1 proteins are used to analyze the effects of ROS inhibition. Altered diffusion and reaction behavior together with the findings of an intramolecular locking of the CRAC channels by $\hzwei$ could explain the reduced CRAC current of ORAI1 channels under $\hzwei$ influence.

The goal of the present study is to transfer these insights of CRAC channel stoichiometries into a new model for CRAC channel formation in order to analyze the different possibilities of ORAI1 channel configurations and their contribution to $\cazweiplus$ influx into the cell.

Therefore we combine dimeric, tetrameric and hexameric structures of ORAI1 channels and adjust the interaction regions to include as many junctional regions as can be observed in the experiment. Furthermore, we account for WT ORAI1 and ROS-preincubated or mutated ORAI1. 

The paper is organized as follows: In section \ref{model} the reaction-diffusion-model is introduced. 
We distinguish two scenarios: the base case scenario and the $\hzwei$ scenario, which deals with a mixture of inhibited and non-inhibited ORAI1 channels. 
Consequently section \ref{Results} deals with the analytic, numeric and stochastic results. Finally, we discuss (\ref{Discussion}) our results.

\section{Model}\label{model}
\begin{figure}
\includegraphics[scale=0.15]{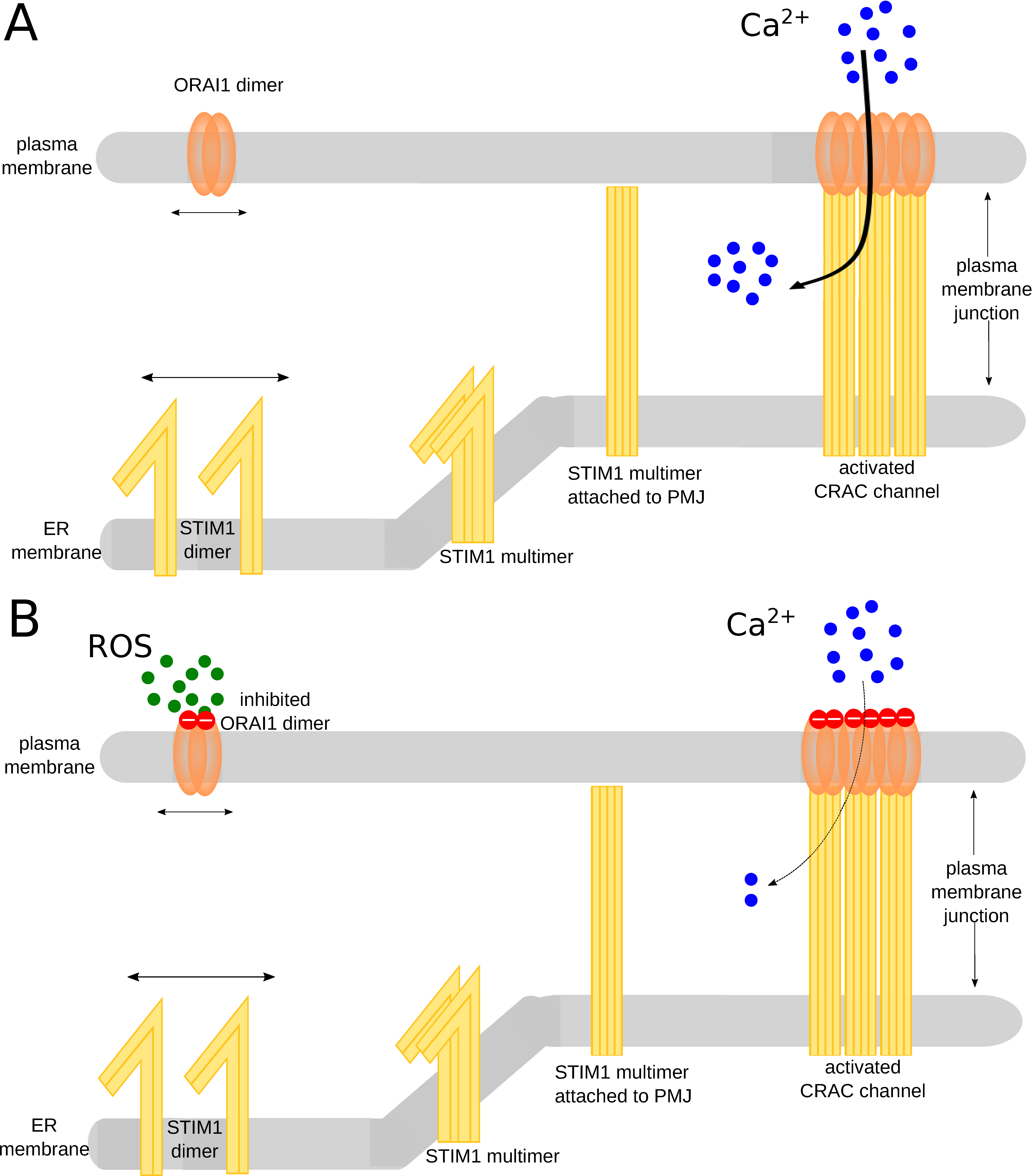}
\caption{\label{fig:ModelBase}\textbf{Schematic representation of CRAC channel formation} \textbf{A:} different steps for CRAC channel formation of the base case scenario, \textbf{B:} different steps for CRAC channel formation under $\hzwei$ pressure, i.e. the $\hzwei$ scenario}
\end{figure}

The formation of CRAC channels requires the interaction of two proteins, STIM1 and ORAI1. The STIM1 proteins are located at the ER membrane preassembled in form of dimers \cite{zhou2015}. Similarly the ORAI1 proteins located at the plasma membrane are assumed as dimers as smallest unit. As soon as the STIM1 EF-hands, which reach into the ER lumen, sense a decrease in the luminal $\cazweiplus$ concentration, the STIM1 proteins unfold and start to multimerize. These STIM1 multimers accumulate at regions near the plasma membrane, the plasma membrane junctions (PMJ). Here they attach to the plasma membrane and trap the ORAI1 proteins. Already a channel consisting of four STIM1 and two ORAI1 (i. e. in our model one CRAC channel subunit) is functional although with a low conductance. Also tetrameric or hexameric CRAC channel structures have been observed \cite{li2016}. These channels consisting of two or three CRAC channel subunits respectively yield an even higher $\cazweiplus$ influx per channel. Figure \ref{fig:ModelBase}A schematically represents the different steps of CRAC channel formation.

An external factor influencing the $\cazweiplus$ influx through CRAC channels is the presence of ROS within the extracellular space. ORAI1 proteins exposed to ROS become inactivated, which means they can still form CRAC channels but the conductance of these inhibited channels is small in comparison to non-inhibited channels or even zero (cf. figure \ref{fig:ModelBase}B). 

Based on the experimental observation described above we now formulate our theoretical model to analyze the ORAI1-STIM1 interaction during CRAC channel formation and the resulting CRAC current.
In the following we assume that the ER store is already depleted, which is experimentally realized by adding 1 $\mu M$ of the SERCA inhibitor thapsigargin. Then we describe CRAC channel formation by the following four steps:
\begin{enumerate}
\item Multimerization of the STIM1 proteins
\item Diffusion of STIM1 and ORAI1 proteins to the PMJ
\item Attachment of STIM1 proteins to a PMJ 
\item Binding of STIM1 and ORAI1 proteins at the PMJ
\end{enumerate}
The multimerization of the STIM1 proteins is given by:
\begin{align}
S_{dimer} + S_{dimer} &\xrightleftharpoons [{k_{demulti}}]{{k_{multi}}} S \, ,\label{eq:StimMultimerize}
\end{align}
where $S_{dimer}$ represents a STIM1 dimer (resting state configuration of STIM1) and $S$ denotes a STIM1 multimer (tetramer), which is able to form a CRAC channel subunit when combined with an ORAI1 dimer. The rates $k_{poly}$ and $k_{depoly}$ define the rates of multimerization and demultimerization respectively. Since after ER store depletion a strong multimerization of STIM1 dimers is observed, one has $k_{multi} \gg k_{demulti}$.
The stationary solution is given by:
\begin{align}
S = \frac{1}{2} \frac{k_{multi}}{k_{demulti}}S_{dimer}^2 \, .\label{eq:StimMulti}
\end{align}

The ORAI1 dimers diffuse within the cell membrane, the STIM1 dimers and tetramers diffuse within the ER plasma membrane. The diffusion rate within the PMJ is assumed to be smaller than outside the plasma membrane junction \cite{wu2014}. The different channel states ($Z_1$, $Z_2$ and $Z_3$) and the STIM1 tetramers connected to the PMJ ($S_{rest}$) are assumed to rest at the PMJ.
Furthermore the reactions
\begin{align}
S + PMJ &\xrightleftharpoons[k_{detach}]{k_{attach}} S_{rest}\label{eq:StimFix}
\end{align}
describe the process of STIM1 tetramers attaching to the PMJ, which means these STIM1 tetrames $S_{rest}$ are fixed at the PMJ and are able to trap an ORAI1 dimer nearby.

The core reaction scheme of CRAC channel formation is given by the following reactions, where $S_{rest}$ denotes the concentration of STIM1 tetramers fixed at the PMJ, O denotes the concentration of ORAI1 dimers within the PMJ and $Z_1, Z_2, Z_3$ denote the concentration of the three different CRAC channel states:
\begin{align}
S_{rest} + O &\xrightleftharpoons [{k_2}]{{k_1}}  Z_1\label{eq:Build1} \\
Z_1 + Z_1 &\xrightleftharpoons [{k_4}]{{k_3}} Z_2\label{eq:Build2}\\
Z_2 + Z_1 &\xrightleftharpoons [{k_6}]{{k_5}} Z_3\label{eq:Build3}
\end{align}
Resting STIM1 tetramers ($S_{rest}$) form together with free ORAI1 dimers ($O$) CRAC channel subunits ($Z_1$). Two of these subunits form an intermediate state ($Z_2$). The fully open CRAC channel ($Z_3$) is built out of an intermediate state and another subunit.

Since CRAC channel formation and disassembly seems to be highly dynamic (\cite{hoover2011} and \cite{alansary2016}) we analyze the influence of an additional reaction between a hexamer CRAC channel state $Z_3$ and a CRAC channel subunit $Z_1$:
\begin{align}
Z_3 + Z_1 \xrightarrow[{}]{{k_7}} 2\cdot Z_2 \label{eq:steal}\, .
\end{align}
This reaction \glqq destroys\grqq\, the fully open CRAC channel configuration and builds two intermediate CRAC channel states with lower $\cazweiplus$ conductivity. Therefore we will call this reaction \glqq stealing mechanism\grqq .

\begin{figure}
\includegraphics[scale=0.25]{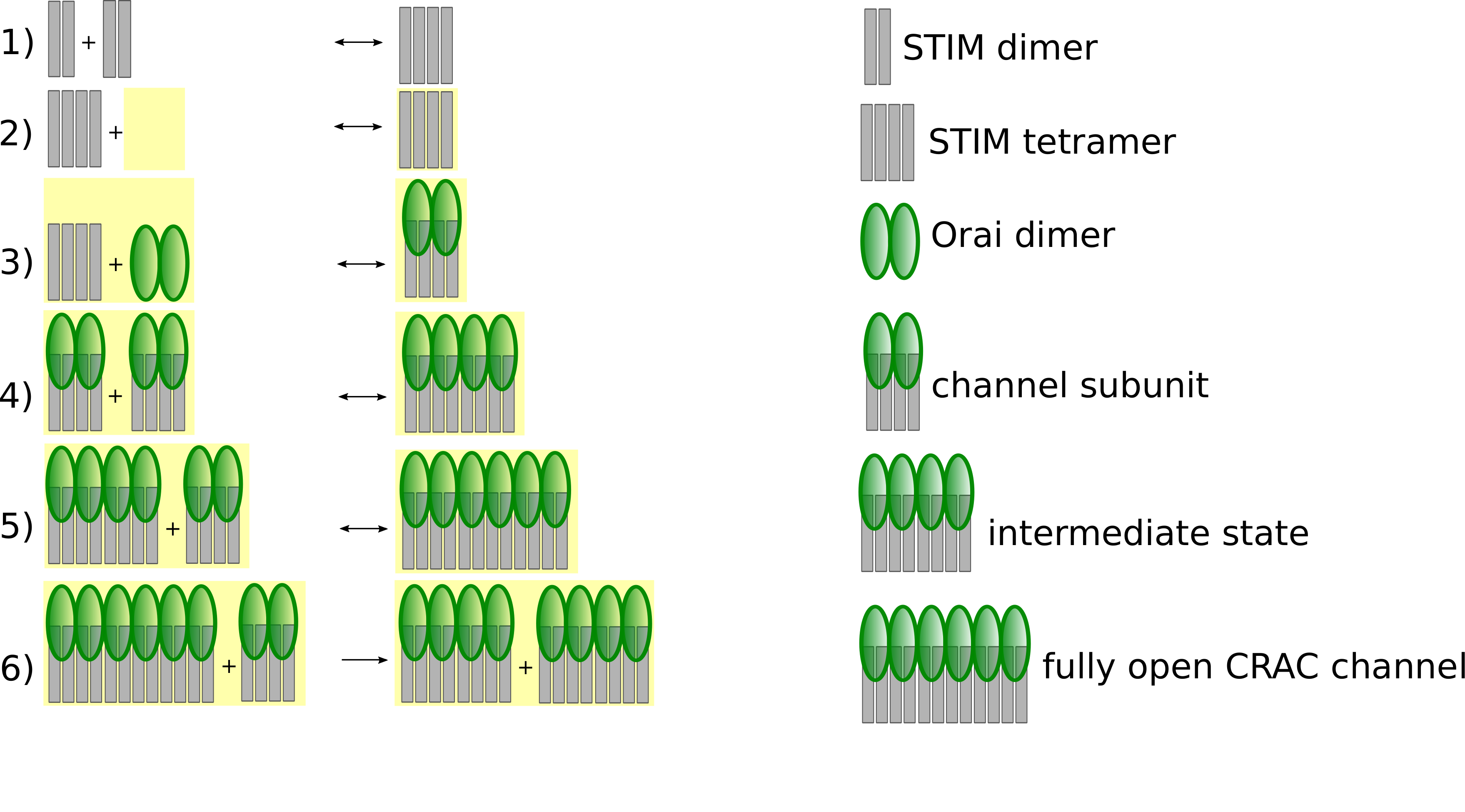}
\caption{\label{fig:BaeCase}\textbf{Schematic representation of the reaction scheme} Yellow boxes indicate reactions, which are allowed at the PMJ regions only.}
\end{figure}

Figure \ref{fig:BaeCase} shows a schematic representation of the reaction scheme. The yellow boxes indicate reactions only allowed at the PMJ. 

The reaction rates $k_1$, $k_3$ and $k_5$ (on-rates) as well as the rates $k_2$, $k_4$ and $k_6$ (off-rates) are not chosen to be independent but are connected through a cooperativity parameter ($\alpha$ respectively $\beta$) to account for the findings of \cite{hoover2011}. In analogy to \cite{peglow2013} the dependence on the rates is assumed to be:
\begin{align}
k_{1+2n} &= \alpha^{n}\cdot k_{1} \, , \, n \in \left(0,1,2\right)\\ \notag
k_{2n} &= \beta^{n}\cdot k_{2} \, , \, n \in \left(0,1,2\right)\, .
\end{align}
As default we examine negative cooperativity ($0 < \alpha, \beta <1$). With this we incorporate the assumption, that it is energetically less favorable to couple more STIM1-ORAI1 complexes to an already established channel, for example due to the confinement of more and more proteins to one single channel complex. This is in line with the findings of [1]. 
Nevertheless within the numerical analysis we also examine positive cooperativity ($\alpha, \beta >1$).
Assuming that the reaction strength between a single CRAC channel subunit and a fully open CRAC channel is as strong as the reaction strength between a CRAC channel subunit and an intermediate state, we take for the rate for the stealing mechanism:
\begin{align}
k_{7} &= k_5 = \alpha^2\cdot k_1 \, .
\end{align}
Because STIM1 multimerization, STIM1 attachment to the PM and STIM1-ORAI1 binding are more preferred than the respective backward reactions, we assume higher rate constants for the forward reactions. Similar to \cite{peglow2013} we assume that STIM1 multimerizing is slower than STIM1-ORAI1 binding. The process of attaching to the membrane is assumed to be very fast, since just one (instead of two) protein complex is involved. Furthermore, we assume strong negative cooperativity.
With the default numerical values for the reaction rate parameters as given in table \ref{table:TabReactionRate} the model provides a reasonable time dependence on $\icrac$.
\begin{table}
\begin{tabular}{c|c}
\textbf{parameter} & \textbf{value} \\ 
\hline 
$k_{multi}$ & $4.8 \cdot 10^5\, l/mol\cdot s$ \\ 
$k_{multi}$ & $0.01 \, 1/s$ \\ 
$k_{attach}$ & $1.8 \cdot 10^6 \, l/mol\cdot s$ \\ 
$k_{detacht}$ & $0.3 \, 1/s$ \\ 
$k_1$ & $1.2 \cdot 10^6 \, l/mol\cdot s$ \\ 
$k_2$ & $1 \, 1/s$ \\ 
$k_3$ & $\alpha \cdot k_1$ \\ 
$k_4$ & $\beta \cdot k_2$ \\ 
$k_5$ & $\alpha^2 \cdot k_1$ \\ 
$k_6$ & $\beta^2 \cdot k_2$ \\ 
$k_7$ & $\alpha^2 \cdot k_1$ \\ 
$\alpha$ & $0.25$ \\
$\beta$ & $0.25$ \\
\end{tabular}
\caption{\textbf{Reaction rate parameters} for the reactions of the basic reaction scheme}
\label{table:TabReactionRate}
\end{table} 

The total $\cazweiplus$ influx into the cell, $\icrac$, is determined by the number of channels $N_{Z_i}$ weighted with their respective conductivity $c_i$:
\begin{align}\label{eq:icracBase}
\icrac = c_1 \cdot N_{Z_1} + c_2 \cdot N_{Z_2} + c_3 \cdot N_{Z_3} \, ,
\end{align}
where we choose $c_1 = 0.01$, $c_2 = 0.25$ and $c_3 = 1.0$.
Experimentally \cite{li2011, dynes2016} it has been observed, that the conductivity of the sub-states is lower than of the fully opened CRAC channel. Furthermore a theoretical analysis of this graded activation was examined in \cite{hoover2011}. With the values of $c_i$ as above we adopt these findings, that the sub-states have lower conductivity than the fully opened CRAC channel. We adjusted the $c_i$ in the way, that our theoretical course of $\icrac(t)$ for $\stot / \otot = 2$ in the ROS scenario agrees with the experimental course of $\icrac(t)$ measured in \cite{alansary2016}. Our values for $c_i$ lay in between the values of the experimental \cite{li2011} and theoretical \cite{hoover2011} findings of prior studies.
The number of channels can be calculated via the formula:
\begin{align}
N_{Z_i} = Z_i \cdot A_{cell} \cdot p_{PMJ} \, ,
\end{align}
where $A_{cell} \approx 530 \mu m^2$ denotes the approximate size of the cell surface area of HEK cells and $p_{PMJ} \approx 15 \%$ describes the fraction of the cell surface where ORAI1 proteins accumulate, the PMJ regions.

In the following we will analyze the reaction-diffusion model as follows:
First, the core reactions without and with stealing mechanism at the PMJ will be analyzed analytically and numerically with special emphasis on the steady state values for the channel states and for $\icrac$.
In a second step we will combine the reaction and diffusion part of the model into a stochastic reaction-diffusion model. Within this model the results of TIRF measurements are used to describe and quantify the changes in PMJ clusters. 

To account for diffusion in the stochastic model the model area is discretized in a grid of $100 \times 100$ grid cells called subvolumes each representing an area of $\Delta l \times \Delta l = 0.1\, \mu m \times 0.1\, \mu m$. The diffusion of the proteins is treated as reactions with reaction rates $d = D/(\Delta l)^2$ assuming periodic boundary conditions. Numerical values for the diffusion constants are given in table \ref{table:TabDiffRate}.

\begin{table}
\begin{tabular}{c|c}
\textbf{diffusion rate} & \textbf{value} \\ 
\hline 
\ \\
$D_O$ & $0.07 \,\mu m^2 / s $ \cite{park2009}\\ 
$D_S$ & $0.1 \,\mu m^2 / s$ \cite{liou2007}\\ 
$D_{S_2}$ & $0.05 \,\mu m^2 / s$ \cite{liou2007}\\ \\
$D_O^{PMJ}$ & $0.03 \,\mu m^2 / s $ \cite{wu2014}\\ 
$D_S^{PMJ}$ & $0.03 \,\mu m^2 / s$ \cite{wu2014}\\ 
$D_{S_2}^{PMJ}$ & $0.015 \,\mu m^2 / s$ (*)\\
\end{tabular}
\caption{\textbf{Diffusion constants} for STIM1 and ORAI1 outside and within the PMJ, (*) assuming that as outside the PMJ the tetramers are half as fast as the dimers}
\label{table:TabDiffRate}
\end{table} 

The two-dimensional model area represents the ER membrane and the plasma membrane simultaneously. The interaction of ORAI1 and STIM1 is only possible at regions, where the ER membrane and the plasma membrane are close together, the plasma membrane junctions PMJ. Therefore, some of the subvolumes of our model area are labeled PMJ. Reaction (\ref{eq:StimMultimerize}) and diffusion is possible in all subvolumes, reactions (\ref{eq:StimFix}) to (\ref{eq:steal}) are just allowed at the PMJ.

\begin{figure}
\includegraphics[scale=0.45]{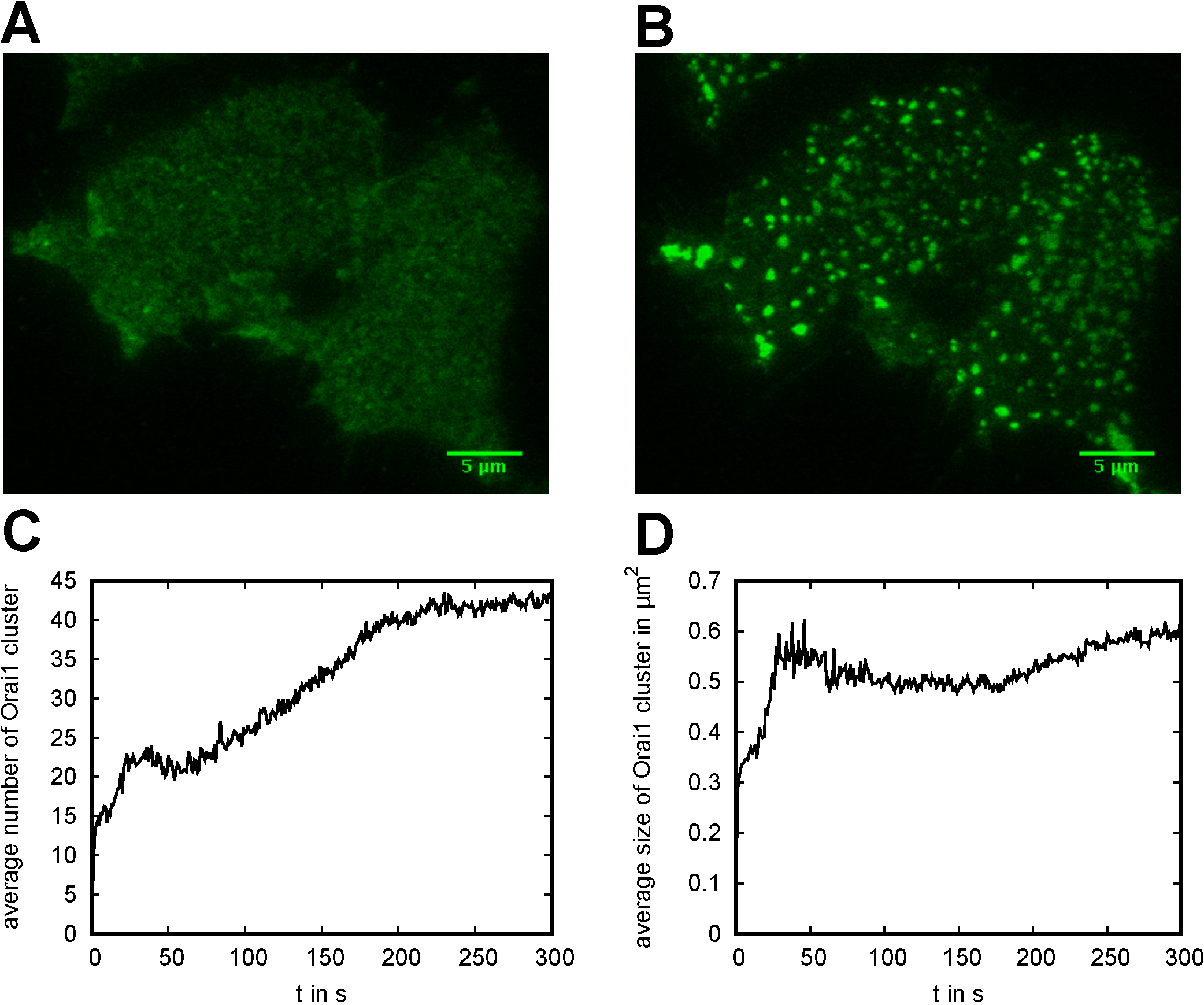}
\caption{\label{fig:TIRF}\textbf{TIRF analysis} \textbf{A:} TRIF images of WT HEKS1 cells with GFP-tagged ORAI1 WT at $t = 0\, s$, \textbf{B:} TRIF images of WT HEKS1 cells with GFP-tagged ORAI1 WT at $t = 300\, s$, \textbf{C:} average number of ORAI1 clusters per cell over time, \textbf{D:} time dependence on the average size of ORAI1 clusters}
\end{figure}

To determine how many subvolumes must be PMJ we analyzed the ORAI1 cluster in HEKS1 cells (i. e. HEK293 stably expressing STIM1 cells) via TIRF microscopy (cf. figure \ref{fig:TIRF}). After addition of the SERCA inhibitor thapsigargin ($1\, \mu M$) leading to passive ER store depletion, the GFP-tagged ORAI1 proteins diffused within the cell membrane to the PMJs. Figure \ref{fig:TIRF}A shows a snapshot of two cells immediately after store depletion, figure \ref{fig:TIRF}B the same cells 300 s later. Analyzing in total 19 cells via TIRF microscopy and with the help of imageJ regarding number and mean size of ORAI1 clusters we find the following (cf. figure \ref{fig:TIRF}C and D): Within the first 50 seconds the number of clusters rises to about 20 per cell and simultaneously the average size of the clusters increases rapidly from $0.3\, \mu m^2$ to about $0.55\, \mu m^2$. After this rapid rise the number of clusters doubles within the next 200 seconds until it reaches a value between 40 and 45. At the same time the average size of clusters drops slightly but then rises again to about $0.6\, \mu m^2$.\\
In our model we increase the number of PMJ subvolumes from one to six within the first 50 seconds and from five to eleven within the next 150 seconds. Simultaneously we increase their size from $0.3\, \mu m^2$ to $0.6\, \mu m^2$ within the first 200 seconds in order to analyze the influence of the PMJ formation. The number of clusters in the model is four times smaller than in the experiment, since the model area is about four times smaller than the area analyzed by TIRF microscopy.

Analogously we analyzed cells preincubated with ROS. Since size and amount of ORAI1 cluster were similar between the two analysis, the same settings in the model are used.

To analyze the reaction-diffusion system described above we perform computer simulations using a Gillespie Monte Carlo algorithm \cite{gillespie1976} with an efficient implementation technique \cite{gibson1999} optimizing the computation time. At the beginning of the simulation ($t=0\, s$) all proteins are homogeneously distributed and the ER $\cazweiplus$ store is assumed to be empty. We start with 30 000 free ORAI1 dimers at the beginning of the simulation. For each case to analyze we average over six runs of simulation.

\subsubsection*{$\hzwei$ scenario}\label{hzweiscenario}
We analyze two different scenarios: The base case scenario described above and the $\hzwei$ scenario to analyze the influence of ROS. To account for the $\hzwei$ preincubated cells we modify the reaction-diffusion scheme as follows:
\begin{itemize}
\item Additional to the CRAC channel states $Z_1$, $Z_2$ and $Z_3$, denoted for better readability as $\one$, $\two$ and $\three$ respectively, composed of STIM1 and ORAI1 proteins, new channel states consisting of STIM1 and preincubated ORAI1 ($\oh$) or of STIM1 and a mixture of $O$ and $\oh$ were included: $\oneh$, $\twoh$, $\twohh$, $\threeh$, $\threehh$, $\threehhh$.
\item New reactions were introduced by exchanging the states $O$, $\one$, $\two$ and $\three$ of reactions (\ref{eq:Build1}) to (\ref{eq:steal}) by the new states in every possible combination.
\item To model $\icrac$ a new parameter $\delta$ was introduced quantifying the degree of inhibition. We assume that the conductance of CRAC channel states consisting of total ROS-inhibited ORAI1 is given by the fraction $\delta$ of the respective CRAC channel state of WT ORAI1. To model the conductance of mixed CRAC channel states (consisting of WT and ROS-inhibited ORAI1), we assume that each included WT dimer contributes to the conductance as before and each included ROS-inhibited dimer contributes with fraction $\delta$ of the WT case. This means that the intermediate state $\Ket{OO^H}$ is weighted with the factor $\left(1 + \delta\right)/2$ and the mixed hexamer CRAC channel states $\Ket{OOO^H}$ and $\Ket{OO^HO^H}$ are weighted with the factors $\left(2 + \delta\right)/3$ and $\left(1 + 2\delta\right)/3$ respectively: 
\begin{align}\label{eq:icracmix}
\icrach &= c_1 \cdot \no + c_2 \cdot \noo + c_3 \cdot \nooo \notag\\
 &+ \delta \left(c_1 \cdot \noh + c_2 \cdot \nohoh + c_3 \cdot \nohohoh\right)\notag\\
 &+ \frac{c_2}{2}\left(1 + \delta\right)\cdot \nooh\notag\\
 &+ \frac{c_3}{3}\left(2 + \delta\right)\cdot \noooh + \frac{c_3}{3}\left(1 + 2\delta\right)\cdot \noohoh
\end{align}
\end{itemize}

In \cite{alansary2016} the interactions between STIM1 and ORAI1 proteins and between ORAI1 and ORAI1 proteins were analyzed. Foerster Resonance Energy Transfer (FRET) values of STIM1 proteins with ORAI1 proteins inhibited by $\hzwei$ were significantly larger ($0.24 \pm 0.03$) than FRET values of STIM1 proteins with WT ORAI1 ($0.15 \pm 0.01$). In contrast, the ORAI1-ORAI1 subunit interaction was reduced by $46\%$.
Fluorescence recovery after photobleaching (FRAP)  measurements revealed a 1.7 times smaller rate of recovery for preincubated ORAI1 proteins \cite{alansary2016}. 
Therefore the diffusion constant of $\oh$ and the reaction rates $\tilde{k}_1$ and $\tilde{k}_{2}$ as well as the cooperativity parameters $\tilde{\alpha}$ and $\tilde{\beta}$ of the new reactions are defined as:
\begin{align}\label{eq:mod_rates}
D_{\oh} &= D_O / 1.7 \\ \notag
\tilde{k}_i &= k_i \cdot 1.6 &\qquad {\rm for} &\qquad i \in (1,2) \\ \notag
\tilde{\alpha} &= \alpha \cdot 0.46  \\ \notag
\tilde{\beta} &= \beta \cdot 0.46\, .
\end{align}

\section{Results}\label{Results}
\subsection{Analytical and numerical analysis of the Base Case Scenario}
Within the analysis of the base case scenario, we examine two reaction systems. The first reaction system consists of the core reactions only, the second reaction system includes the additional stealing mechanism. The core reaction system reads:

\begin{align}\label{reactionsystem_basic}
\frac{dS}{dt}  &= -k_1\cdot S \cdot O + k_2 \cdot Z_1\\ \notag
\frac{dO}{dt}  &= -k_1\cdot S \cdot O + k_2 \cdot Z_1\\ \notag
\frac{dZ_1}{dt} &= k_1\cdot S \cdot O - k_2 \cdot Z_1 - k_3\cdot \frac{Z_1^2}{2}\cdot 2 + 2\cdot k_4\cdot Z_2\\ \notag
& -k_5\cdot Z_2 \cdot Z_1 + k_6 \cdot Z_3\\ \notag
\frac{dZ_2}{dt} &=
k_3\cdot \frac{Z_1^2}{2} -k_4 \cdot Z_2 -k_5 \cdot Z_2\cdot Z_1 + k_6 \cdot Z_3 \\\notag
\frac{dZ_3}{dt} &=k_5\cdot Z_2 \cdot Z_1 - k_6\cdot Z_3 \notag
\end{align}
and when including the stealing mechanism, the right hand sides of (\ref{reactionsystem_basic}) are extended by:

\begin{align}\label{eq:stealing_reaction_scheme}
\frac{dZ_{1,3}}{dt} \to&\, \frac{dZ_{1,3}}{dt} -k_7\cdot Z_1\cdot Z_3 \\\notag
\frac{dZ_2}{dt} \to&\, \frac{dZ_2}{dt} +2\cdot k_7\cdot Z_1\cdot Z_3\, . \\\notag
\end{align}

The total amounts of ORAI1 (in whichever CRAC channel configuration) and of STIM1 respectively have to be conserved, which means:
\begin{align}\label{additional_condition_O}
O^{tot} = O(t) + Z_1(t) + 2\cdot Z_2(t) + 3\cdot Z_3(t)\, ,
\end{align}
\begin{align}\label{additional_condition_S}
S^{tot} = S(t) + Z_1(t) + 2\cdot Z_2(t) + 3\cdot Z_3(t)\, .
\end{align}
The second condition is always fulfilled when (\ref{additional_condition_O}) is fulfilled, since the right hand sides of the differential equations for $O$ and $S$ are identical and we consider the following initial conditions:
\begin{align}\label{eq:initial}
S(0) &= S^{tot} \\ \notag
O(0) &= O^{tot} \\ \notag
Z_1(0) &= 0 \\ \notag
Z_2(0) &= 0 \\ \notag
Z_3(0) &= 0 \, .
\end{align}

Considering the reaction system without stealing (eqns (\ref{reactionsystem_basic})), the second equation of (\ref{reactionsystem_basic}) can be rewritten as:
\begin{align}
\frac{dZ_1}{dt} &= -\left(\frac{dO}{dt} + 2\cdot\frac{dZ_2}{dt} + 3\cdot\frac{dZ_3}{dt}\right) \, ,
\end{align}
which results in 
\begin{align}
Z_1(t) = -\left(O(t) + 2\cdot Z_2(t) + 3\cdot Z_3(t) + c_0\right) \, ,
\end{align}
where $c_0$ is determined by the initial condition (\ref{eq:initial}), which yields $c_0 = O^{tot}$.
This is identical to the additional condition above (\ref{additional_condition_O}).

The stationary state of the reaction system (\ref{reactionsystem_basic}) obeys:
\begin{align}
0 =& -k_1\cdot S^{stat} \cdot O^{stat} + k_2 \cdot Z^{stat}_1 \\ \notag
0 =&\, k_3\cdot \frac{{Z^{stat}_1}^2}{2} -k_4 \cdot Z^{stat}_2 -k_5 \cdot Z^{stat}_2\cdot Z^{stat}_1\\ \notag
& + k_6 \cdot Z^{stat}_3\\ \notag
0 =&\, k_5\cdot Z^{stat}_2 \cdot Z^{stat}_1 - k_6\cdot Z^{stat}_3 \, . \notag
\end{align}
These are three independent equations with five concentration parameters ($S^{stat},\,O^{stat},\,Z^{stat}_1,\,Z^{stat}_2,\,Z^{stat}_3,\,$) and six reaction rates $k_i$.

With the cooperativity parameters $\alpha$ and $\beta$ for the on- and off-rates and using the abbreviations
\begin{align}
\gamma &= \frac{\alpha}{\beta}\\ \notag
k &= \frac{k_1}{k_2} \, ,
\end{align}
the equation system can be rewritten in terms of the stationary number of free ORAI1 and free STIM1 proteins:
\begin{align}\label{eq:steady_state_basic}
Z^{stat}_1 &= k\cdot S^{stat}\cdot O^{stat}\\ \notag
Z^{stat}_2 &= 0.5\cdot \gamma\cdot k\cdot S^{{stat}^2} \cdot O^{{stat}^2}\\ \notag
Z^{stat}_3 &= 0.5\cdot \gamma^3 \cdot k^2 \cdot S^{{stat}^3}\cdot O^{{stat}^3} \, .
\end{align}
Together with the conservation laws (\ref{additional_condition_O}) and (\ref{additional_condition_S}) one gets a polynomial equation for $Z^{stat}_1$, which can be solved numerically. All other stationary concentrations then follow.

Similarly we analyzed the reaction system with stealing mechanism (eqns (\ref{eq:stealing_reaction_scheme})):
The additional stealing reaction (\ref{eq:steal}) should help to regulate the formation of CRAC channels.
The corresponding stationary equation system together with the two conservation laws (\ref{additional_condition_O}) and (\ref{additional_condition_S}) can be simplified to the following non-linear, non-homogeneous equation system:
\begin{align}
O^{ges} =&\, O^{stat} + Z_1^{stat} + 2\cdot Z_2^{stat} + 3\cdot Z_3^{stat} \\ \notag
S^{ges} =&\, S^{stat} + Z_1^{stat} + 2\cdot Z_2^{stat} + 3\cdot Z_3^{stat} \\ \notag
Z_1^{stat} =&\, \frac{k_1}{k_2}S^{stat}\cdot O^{stat}\\ \notag
Z_2^{stat} =&\, \frac{k_3}{2k_4}(Z_1^{stat})^2 + \frac{k_7}{k_4}Z_1^{stat}\cdot Z_3^{stat}\\ \notag
Z_3^{stat} =&\, \frac{k_5\cdot Z_1^{stat}\cdot Z_2^{stat}}{k_6 + k_7Z_1^{stat}}\, ,
\end{align}
which is consistent with the basic solution for $k_7 = 0$. 

\subsubsection*{Time evolution of the CRAC channel states}
\begin{figure}
\includegraphics[scale=0.75]{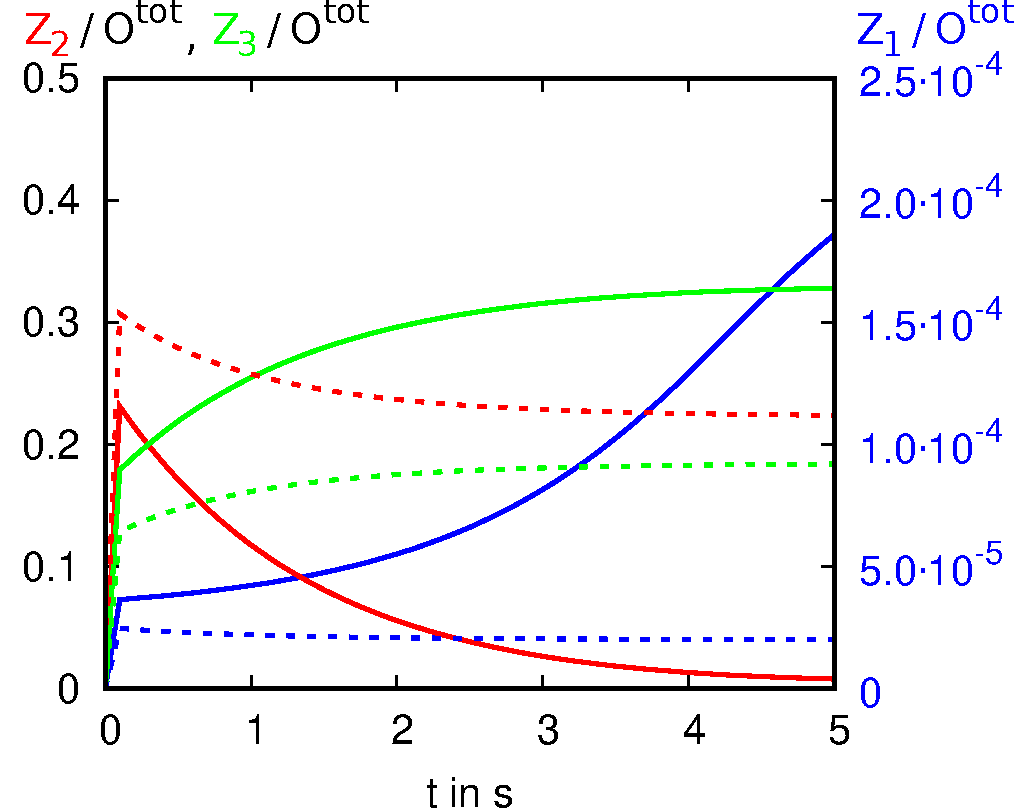}
\caption{\label{fig:basic}\textbf{Time evolution of CRAC channel states} $Z_1$ (blue), $Z_2$ (red) and $Z_3$ (green) according to the core reaction scheme (eq (\ref{reactionsystem_basic}),  solid lines) and the reaction system with stealing mechanism (eq (\ref{eq:stealing_reaction_scheme}), dashed lines) for the rate parameters as given in table \ref{table:TabReactionRate} and for $O^{tot} = 120000$ and $S^{tot} = 240000$ for $t \in \left[0\, s, 5\, s\right]$.}
\end{figure}

The reaction systems (\ref{reactionsystem_basic}) and (\ref{eq:stealing_reaction_scheme}) can be solved numerically. 
Figure \ref{fig:basic} exemplifies the time evolution of all three CRAC channel states normalized to the total amount of ORAI1 ($\otot$) within the first five seconds, with the parameters given in table \ref{table:TabReactionRate} and for $O^{tot} = 120000$ and $S^{tot} = 240000$ for $t \in \left[0\, s, 5\, s\right]$. 
We estimated the total number of ORAI1 dimers as follows: The area of one cluster is about $0.6 \,\mu m^2$ large according to the TIRF measurements. This corresponds to a square with side length of about 775 nm. We assume that the ORAI hexamer has a diameter of 10 nm and that we can place 60x60 = 3600 ORAI hexamers into one cluster. These corresponds to 10 800 ORAI dimers. In the stochastic simulations we will have eleven clusters of size $0.6 \,\mu m^2$ and therefore 11x10800 = 118 800 ORAI dimers. Due to this rough estimate we set the numbers of ORAI1 within the numeric calculation to 120 000. To operate with twice the number of STIM1 than ORAI1, we set the number of STIM1 to 240 000.
The solid lines represent the core reaction system (\ref{reactionsystem_basic}) and the dashed lines represent the system with stealing mechanism (\ref{eq:stealing_reaction_scheme}).
In both cases the single subunit state $Z_1$ is rarely occupied (less than $2 \cdot 10^{-4}$), as on-reactions are more preferred than off-reactions, while $Z_2$ and $Z_3$ show more interesting features:

The time course of $Z_2$ and $Z_3$ is determined by the ratio of $k_1$ and $k_2$. 
Analyzing the base case scenario without stealing mechanism shows the following: The larger $k_1 / k_2$ the steeper the rise of $Z_2$ and $Z_3$ within the first millisecond, the steeper the following rise of $Z_3$ within the next 50 seconds and the steeper the fall of $Z_2$ within this time.
Analysis of the base case scenario with stealing mechanism shows similar dependence on the time evolution of $Z_2$ and $Z_3$. The main differences occur in the absolute values of these two states the tetrameric state is more occupied than the hexameric state.
In experiments (\cite{alansary2016}) it is observed that the delay between tapsigargin addition and the steady state $\cazweiplus$  influx is larger than two seconds, which is the case for the base case scenario with default values (cf. figure  \ref{fig:basic}). This means that the delay is also influenced by the diffusion of STIM1 and ORAI1 proteins towards the PMJ. 
\subsubsection*{Influence of the rate parameters}
\begin{figure}
\includegraphics[scale=0.5]{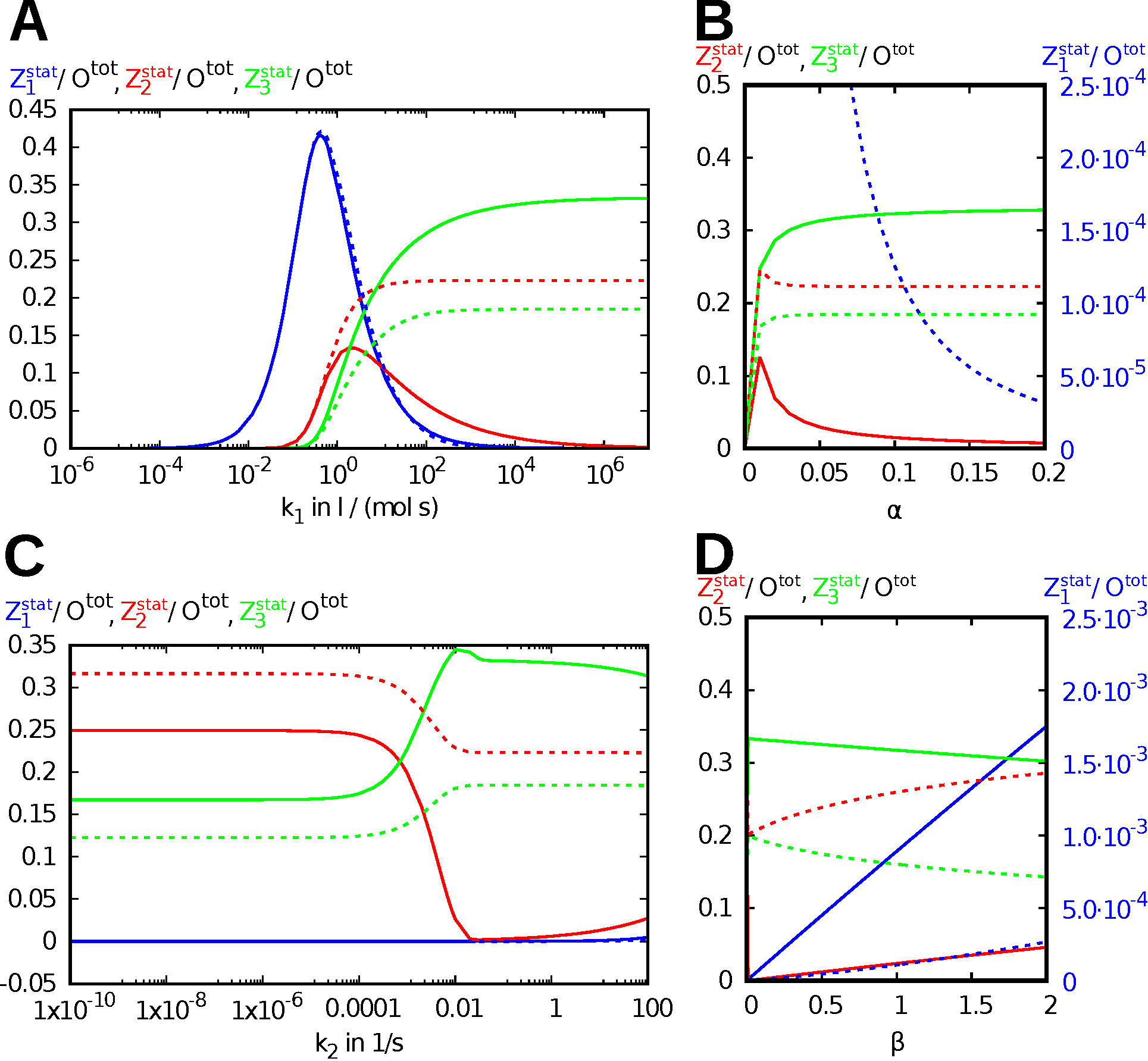}
\caption{\label{fig:rate_numeric}\textbf{$\icrac$ in dependence on different rate parameters} Graphic representation of the values of the steady state configuration (i. e. at $t = 300\, s$) $Z^{stat}_1$ (blue), $Z^{stat}_2$ (red) and $Z^{stat}_3$ (green) in dependence on \textbf{A:}  $k_1$, \textbf{B:} $\alpha$, \textbf{C:}  $k_2$ and \textbf{D:} $\beta$. The core reaction system (eq (\ref{reactionsystem_basic})) is represented by solid lines, the reaction system with stealing mechanism (eq (\ref{eq:stealing_reaction_scheme})) is represented by dashed lines. If not varied as displayed on the x-axis, the rate parameters are chosen according to table \ref{table:TabReactionRate}.}
\end{figure}
In a second step we analyzed the influence of the rate parameters ($k_1$, $k_2$, $\alpha$ and $\beta$) on the steady state values. Keeping all but the parameters displayed on the x-axis fixed as given in table \ref{table:TabReactionRate}, figure \ref{fig:rate_numeric} shows the results for varying the respective parameters. Solid lines indicate the core reaction system, while dashed lines indicate the inclusion of the stealing reaction.

Figure \ref{fig:rate_numeric}A shows the influence of the on-rate $k_1$ on the stationary solution of the reaction schemes. In both cases the behavior of $Z_1$ is similar: Only in the case when the on-rate is small (between $10^{-2}\, l/(mol\cdot s)$ and $10^{2}\, l/(mol\cdot s)$) a lot of dimeric CRAC channel states can be found. If the on-rate is even less than $10^{-3}\, l/(mol\cdot s)$ no channels at all are formed. The behavior of intermediate CRAC channel states on changes of $k_1$ differs without (\ref{reactionsystem_basic}) or with (\ref{eq:stealing_reaction_scheme}) stealing mechanism. In the second case a larger on-rate yields to more tetrameric CRAC channel states up to the value 0.22. In the first case just for on-rates between $10^{-1} \, l/(mol\cdot s)$ to $10^{5} \, l/(mol\cdot s)$ a non negligible stationary value of $Z_2$ is reached. As already expected from the data of the time evolution analysis, the most dominant state of the core reaction system is the hexamer configuration for large on-rates. In case of core reactions with stealing mechanism the stationary value of $Z_3$ is smaller than the one of $Z_2$ for large on-rates, which means, that more tetramers than hexamers are formed.

Keeping the on-rate fixed and varying the off-rate (figure \ref{fig:rate_numeric}C) results always in no noticeable values of $Z_1$. This is obvious from the data before, because in the examined interval the on-reactions are always more preferred than the off-reactions. In the core reaction scenario the tetramer configuration is more often adopted as the hexamer configuration as long as the off-rate is smaller than $10^{-3}\, 1/s$ and then drops to almost zero for larger off-rates. In contrast to this the hexamer configuration starts at a value of 0.177 for small off-rates, rises to a value of 0.344 at $k_2 = 0.01$ and than slightly drops again to 0.289 for even larger off-rates. Regarding the reaction system with stealing mechanism the tetramer configuration is always favored, while the gap between tetramer and hexamer configuration scales down from 0.2 for off-rates between $10^{-9}$ to $10^{-3}$ to 0.04 for larger off-rates.\\
Figure \ref{fig:rate_numeric}B shows the dependence on the CRAC channel states on the on-rate cooperativity parameter $\alpha$. Only for small $\alpha$ (smaller than 0.05) compared to the off-rate cooperativity parameter $\beta$ (0.25) the steady state values of $Z_2$ and $Z_3$ develop from zero to their respective constant value as in the analysis above. $Z_1$ drops from 1 ($\alpha = 0$) to $10^{-3}$ ($\alpha = 0.1$) and than to less than $10^{-5}$ for $\alpha$ rising until 0.2.

In contrast to this $Z_i$ change more significantly on changes of the off-rate cooperativity parameter $\beta$, when $\alpha$ is fixed at 0.25 (cf. figure \ref{fig:rate_numeric}D). For the core reaction system $Z_1$ rises linear from zero to $2.5\cdot 10^{-4}$ for $\beta$ between zero and two. Similarly $Z_2$ rises from zero to 0.29 and $Z_3$ drops from 0.2 to 0.14 for $\beta$ between $0.1$ and 2.
\subsubsection*{Dependence of $\icrac$ on the ratio $S^{ges}/O^{ges}$}
\begin{figure}
\includegraphics[scale=0.6]{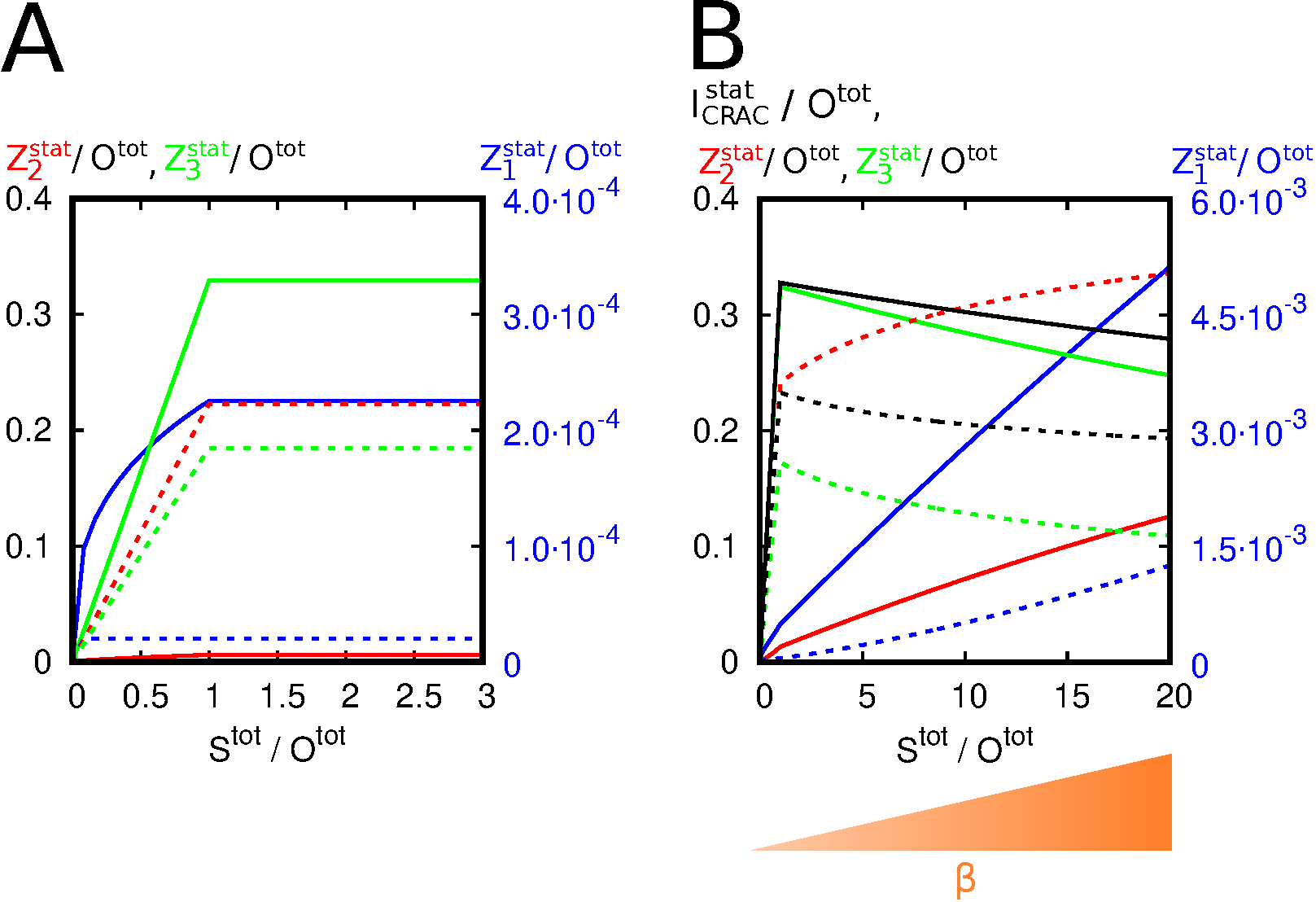}
\caption{\label{fig:svso_numeric} \textbf{$\icrac$ in dependence on the ratio STIM1 / ORAI1} Graphic representation of the values of the steady state configuration $Z^{stat}_1$ (blue, right y-axis), $Z^{stat}_2$ (red) and $Z^{stat}_3$ (green) \textbf{A:} in dependence on the ratio STIM1 / ORAI1, \textbf{B: } in dependence on the ratio STIM1 / ORAI1 with additional increase of $\beta$ (unbinding cooperativity parameter).
Solid lines indicate the core reaction scheme (eq (\ref{reactionsystem_basic})), dashed lines show the corresponding values for the reaction scheme including the stealing mechanism (eq (\ref{eq:stealing_reaction_scheme})). Parameters are chosen as given in table \ref{table:TabReactionRate} and in \textbf{B} $\beta$ is varied linear in the interval $[0.25, 6.25]$.}
\end{figure}
Major parameters influencing $\icrac$ are the amount and ratio of STIM1 and ORAI1 proteins. Figure \ref{fig:svso_numeric}A shows the results for the different stationary CRAC channel states normalized to $\otot$. As soon as there are twice or more as many STIM1 dimers than ORAI1 dimers the stationary values of $Z_i$ are constant. For ratios of STIM1 to ORAI1 below two we see a linear rise of the steady state values $Z_{2,3}$ form zero to their respective value at STIM1 / ORAI1 = 2 in both cases.  $Z_1$ of the core reaction scheme first shows a steep linear and than slightly slower rise to its final value.
In comparison, figure \ref{fig:svso_numeric}B shows the stationary values of $Z_i$ and $\icrac$ normalized to $\otot$ in dependence on the ratio $S^{tot}/O^{tot}$ with simultaneous rise of $\beta$ (linear from 0.25 to 6.25).
This scenario corresponds to the assumption that it gets more difficult for CRAC channel units to build up larger complexes as more and more STIM1 proteins arrive at the PMJ.
In this case all $Z_i$ first rise between $S^{tot}/O^{tot} = 0$ and $S^{tot}/O^{tot} = 2$ ($0.25 \le \beta \le 0.55$). Subsequently, $Z_1$ and $Z_2$ rise more slowly whereas $Z_3$ drops for larger $\beta$ and larger ratios $S^{tot}/O^{tot}$. This results in both reaction systems (without stealing (\ref{reactionsystem_basic}), solid line, and with stealing (\ref{eq:stealing_reaction_scheme}), dashed line) in a slight decrease of $\icrac$. 
A similar behavior is seen experimentally in \cite{kilch2013}.
\subsection{Stochastic analysis of the Base Case Scenaraio}
\subsubsection*{Analysis of $\icrac$ in dependence on time and of the ratio STIM1 / ORAI1}
Main aspects of the analysis of the base case scenario are the dependence on $\icrac$ of time t and of the ratio of STIM1 proteins to ORAI1 proteins.
\begin{figure}
\includegraphics[scale=0.45]{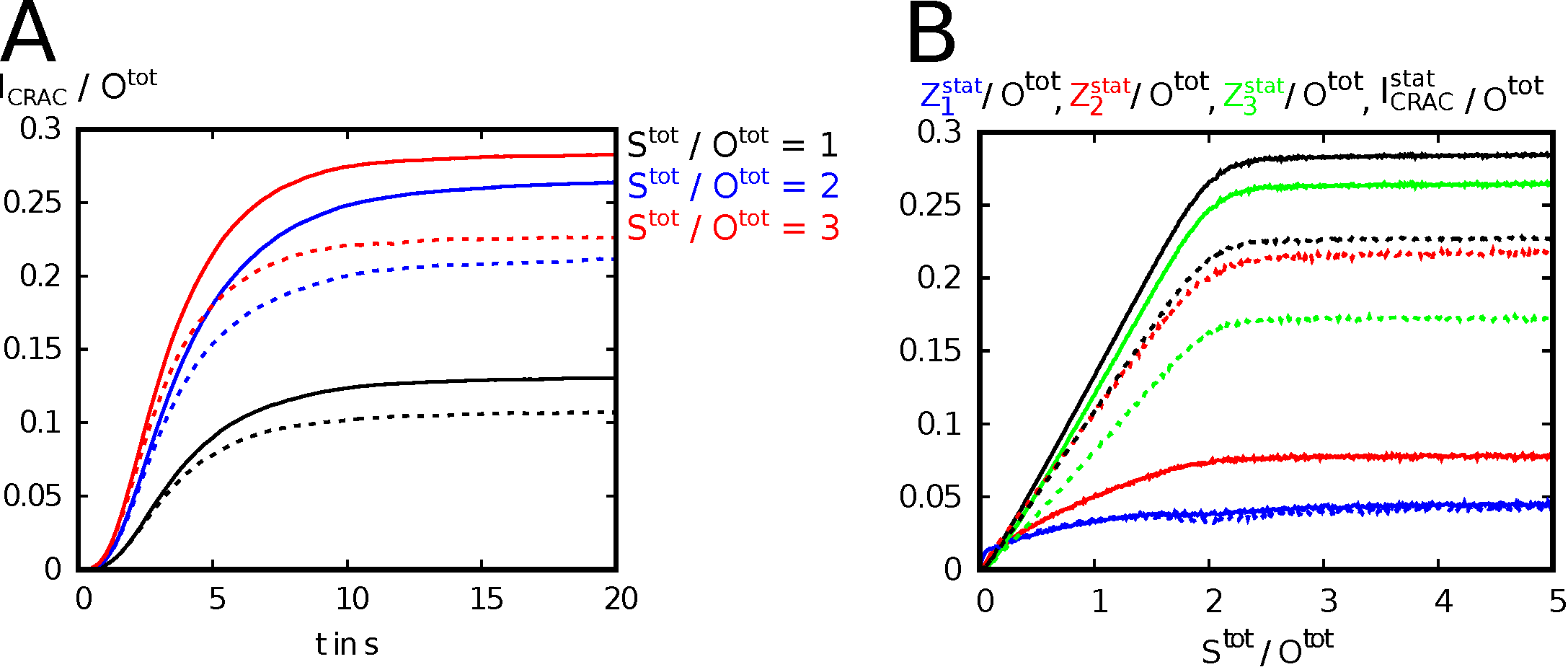}
\caption{\label{fig:ICRACstochastic}\textbf{Base case scenario with different ratios $\stot / \otot$} \textbf{A:} temporal course of $\icrac$ for different ratios of STIM1 / ORAI1, \textbf{B:} steady state values (i. e. at $t = 300$) of CRAC channel states $Z_1^{stat}$, $Z_2^{stat}$, $Z_3^{stat}$ (blue, red, green respectively) and  of $\icracstat$ (black) in dependence on the ratio STIM1 / ORAI1. Solid lines indicate the core reaction scheme (eq (\ref{reactionsystem_basic})), dashed lines show the corresponding values for the reaction scheme including the stealing mechanism (eq (\ref{eq:stealing_reaction_scheme})). We average over six simulations, the errorbars are smaller than the datapoints.}
\end{figure}
Figure \ref{fig:ICRACstochastic}A shows the development of $\icrac$ normalized to the total available amount of free ORAI1 dimers within the first 20 seconds for three different ratios STIM1 / ORAI1 for the core reaction scheme without (solid lines) and with (dashed lines) stealing mechanism.
In contrast to the numerical analysis above the respective steady state value of $\icrac$ is reached after about eight to ten seconds not yet after about two seconds. This is because diffusion of ORAI1 and STIM1 proteins is now taken into account. 
For all three ratios of STIM1 / ORAI1 (STIM1 / ORAI1 = 1, 2 and 3 in black, blue and red respectively) the core reaction scheme without stealing mechanism leads to a higher amount of $\cazweiplus$ influx compared to the reaction scheme with stealing mechanism. In both reaction schemes, $\cazweiplus$ influx is highest for STIM1 / ORAI1 = 3. It is slightly lower for STIM1 / ORAI1 = 2 and significantly lower for STIM1 / ORAI1 = 1.

Figure \ref{fig:ICRACstochastic}B shows the steady state values of $Z^{stat}_i$ and $\icracstat$ normalized to $\otot$ in dependence on the ratio STIM1 / ORAI1. For both reaction schemes without and with stealing the numbers of tetramer and hexamer CRAC channel states rise linear from STIM1 / ORAI1 = 0 to STIM1 / ORAI1 $\simeq$ 2.5 and then stay at a constant value except for statistical fluctuations. The number of CRAC channel subunits $Z_1$ is low in comparison to the number of higher CRAC channel states.

As in the numerical analysis the hexamer states are the most dominant for the base case scenario without stealing mechanism and the tetramer states are the most dominant for the base case scenario with stealing mechanism. This results in lower $\icrac$ in the latter scenario. 

In contrast to the numerical analysis above, the highest steady state value of $\icracstat$ is reached at STIM1 / ORAI1 $\gtrsim$ 2.5 and not at exactly STIM1 / ORAI1 = 2. The latter ratio is the minimum number of STIM1 tetramers needed to saturate all free ORAI1 dimers available and therefore is the theoretical minimal ratio at which the highest value of $\icrac$ is reached. The stochastic analysis includes diffusion of STIM1 and ORAI1. In this case there is more than the minimum amount of STIM1 needed to saturate all ORAI1 dimers, because it is possible, that free STIM1 tetramers and free ORAI1 dimers are not at the same spot and therefore can not react with each other. Are there about $25 \%$ more STIM1 proteins available than the minimal theoretical amount needed the maximal $\cazweiplus$ influx is reached. 

In the following we perfom all analysis with the reactions of the core system with stealing mechanism.

\subsubsection*{Influence of the diffusion and rate constants}
\begin{figure}
\includegraphics[scale=0.4]{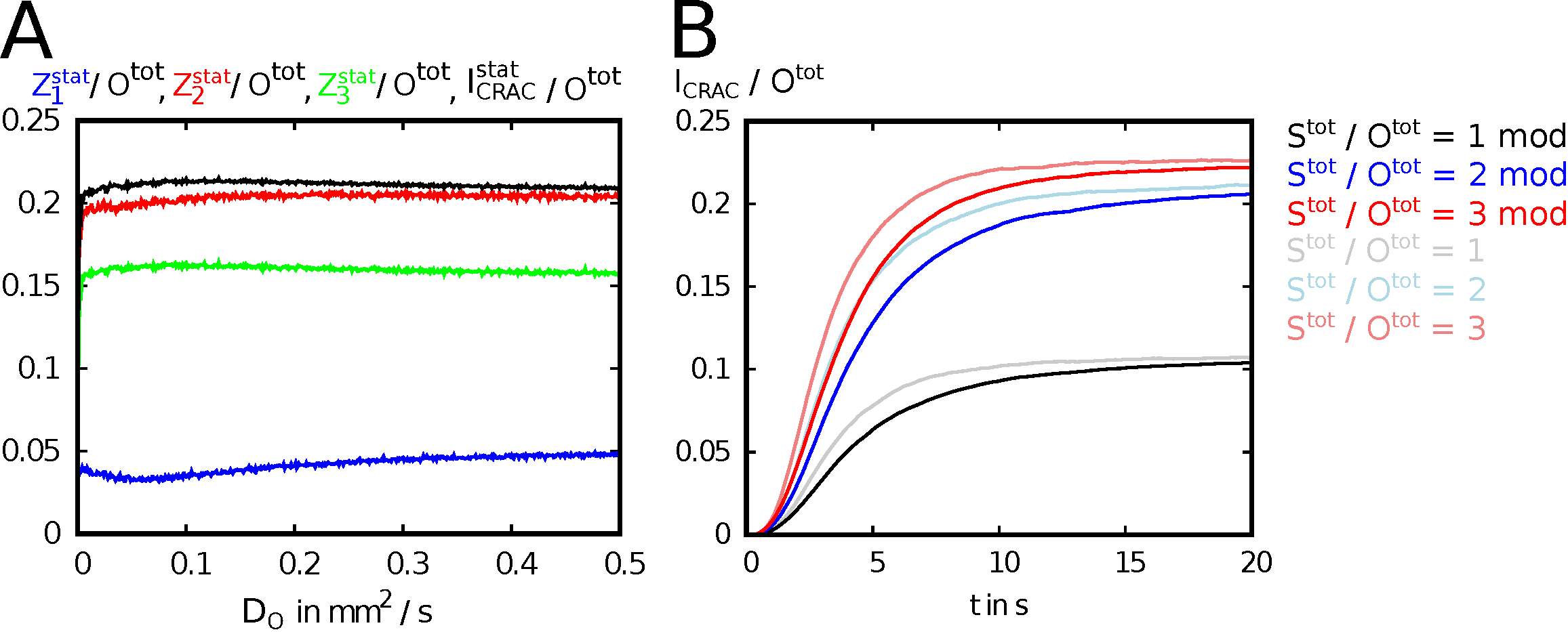}
\caption{\label{fig:Diff}\textbf{Modifying diffusion and rate constants in the base case scenario}\\\textbf{A:} Stochastic analysis of the steady state values (i. e. at $t = 300$) of CRAC channel states $Z_1^{stat}$, $Z_2^{stat}$, $Z_3^{stat}$ (blue, red, green respectively) and  of $\icracstat$ (black) in dependence on the diffusion constant $D_O$ of the ORAI1 dimers according to the reaction scheme including the stealing mechanism (eq (\ref{eq:stealing_reaction_scheme})). \textbf{B:} Temporal course of $\icrac$ according to eq (\ref{eq:stealing_reaction_scheme}). Faded lines indicate the parameters as given in tables \ref{table:TabReactionRate} and \ref{table:TabDiffRate}, bold lines indicate the modified parameters according to eq. (\ref{eq:mod_rates}) and $\tilde{D_O} = D_O/1.7$.  We average over six simulations, the errorbars are smaller than the datapoints.}
\end{figure}
Comparing WT ORAI1 channels with ROS preincubated ORAI1 channels one experimentally measured parameter concerns the diffusion constant of ORAI1. WT ORAI1 proteins are faster than ROS preincubated ORAI1 channels \cite{alansary2016}. To analyze, whether this could influence the total $\cazweiplus$ influx into the cell, we altered the diffusion rate of the ORAI1 dimers outside and inside the PMJ areas in our model. 
Figure \ref{fig:Diff}A shows the amplitude of $\icracstat$ as well as the number of active channels at steady state (i. e. at $t=300\, s$) normalized to $\otot$ in dependence on the diffusion constant of the ORAI1 dimers.
The diffusion constant $D_O$ outside the PMJ regions was varied between 0 and 0.5 $mm^2 / s$. To account for the slower diffusion of ORAI1 within the PMJ, $D^{PMJ}_O$ was set to $3\cdot D_O / 7$ in analogy to the difference in the measured values of \cite{park2009} and \cite{wu2014} (compare table \ref{table:TabDiffRate}). Apart from statistical fluctuations the number of active channels in the steady state and hence the value of $\icracstat$ do not vary for different values of $D_O$.
Consequently, the lowered diffusion constant alone can not explain the drastic decrease of $\cazweiplus$ influx seen in experiments (\cite{alansary2016}) between the $\hzwei$ scenario and the WT scenario.

In a next step in \cite{alansary2016} the interactions between STIM1 and ORAI1 proteins and between ORAI1 and ORAI1 proteins were analyzed. According to this we changed the reaction parameters as described in equation (\ref{eq:mod_rates}).
Combined with the  modified diffusion constant ($\tilde{D_O} = D_O/1.7$) the temporal course of $\icrac$ was examined. The results are shown in figure \ref{fig:Diff}B.
The faded black, blue and red curves show the courses of WT ORAI1 for the three ratios of STIM1 / ORAI1 = 1, 2 and 3 respectively. These courses are being compared to the modified version (mod) of the base case scenario including the altered reaction rates and the altered diffusion constant (black, blue and red respectively).
The differences between the basic reaction scheme and the modified version are small.
\ \\
To sum up, the analysis of the base case scenario leads to the following conclusions:
\begin{itemize}
\item $\icrac$ strongly depends on the ratio of STIM1 / ORAI1. $\cazweiplus$ influx is highest for STIM1 / ORAI1 $\gtrsim$ 2.5.
\item The dominant channel form of the steady state value of $\icracstat$ (at t = 300 s) depends on the possibility of single CRAC channel subunits to disrupt hexameric channels (what we call \glqq stealing mechanism\grqq). Are the subunits not able to disrupt the fully open CRAC channels (core reactions, eqns. (\ref{reactionsystem_basic})) the hexameric state is the most favored state. Including the stealing mechanism eq.(\ref{eq:stealing_reaction_scheme}) leads to a lower number of hexamers and a higher number of tetramers. The total amount of $\cazweiplus$ influx is decreased in this case.
\item Altering the diffusion rate of free ORAI1 dimers does not result in significant changes of $\icracstat$. 
\item Altered reaction and diffusion rates according to \cite{alansary2016} (eq. (\ref{eq:mod_rates}) and $\tilde{D_O} = D_O/1.7$) do not significantly change the temporal course of $\icrac$ for different ratios of STIM1 / ORAI1.
\end{itemize}

\subsection{Stochastic analysis of the $\hzwei$ Scenario} \label{ssec:ResHzwei}
In the next step we extended our model taking the findings of an intramolecular locking of ORAI1 channels by $\hzwei$ (\cite{alansary2016}) into account. Experimentally this corresponds to ORAI1 preincubated by ROS or ORAI1 mutates as described in \cite{alansary2016}. We introduce a new particle species $\oh$, which represents ORAI1 proteins that reacted with ROS (cf. section \ref{hzweiscenario}). Thus three parameters are of major interest:
\begin{itemize}
\item The ratio of the total amount of STIM1 proteins to the total amount of ORAI1 proteins (i. e. free $O$ and $\oh$ as well as all ORAI1 proteins in whichever CRAC channel configuration) $\bm{\stot / \otot}$,
\item the ratio of ROS-inhibited ORAI1 (free and bound in CRAC channels) to the total amount of ORAI1 $\bm{\oh / \otot}$ and
\item the inhibition parameter $\bm{\delta}$ (cf. eq. (\ref{eq:icracmix})).
\end{itemize}
As default values we choose $\stot / \otot = 2$, $\oh / \otot = 0.95$ and $\delta = 0.1$.

To count the different channel states we summed up all different combinations occurring of each type of CRAC channel state, which means
\begin{align}
Z_1 &= Z_1^{0} + Z_1^{H}\, ,\\ \notag
Z_2 &= Z_2^{0} + Z_2^{H} + Z_2^{2H}\, \text{and}\\ \notag
Z_3 &= Z_3^{0} + Z_3^{H} + Z_3^{2H} + Z_3^{3H}\, 
\end{align}
where the numbers of $H$ in the exponent indicate how many ORAI1 dimers that interacted with ROS are part of the respective CRAC channel. 

\begin{figure}
\includegraphics[scale=0.3]{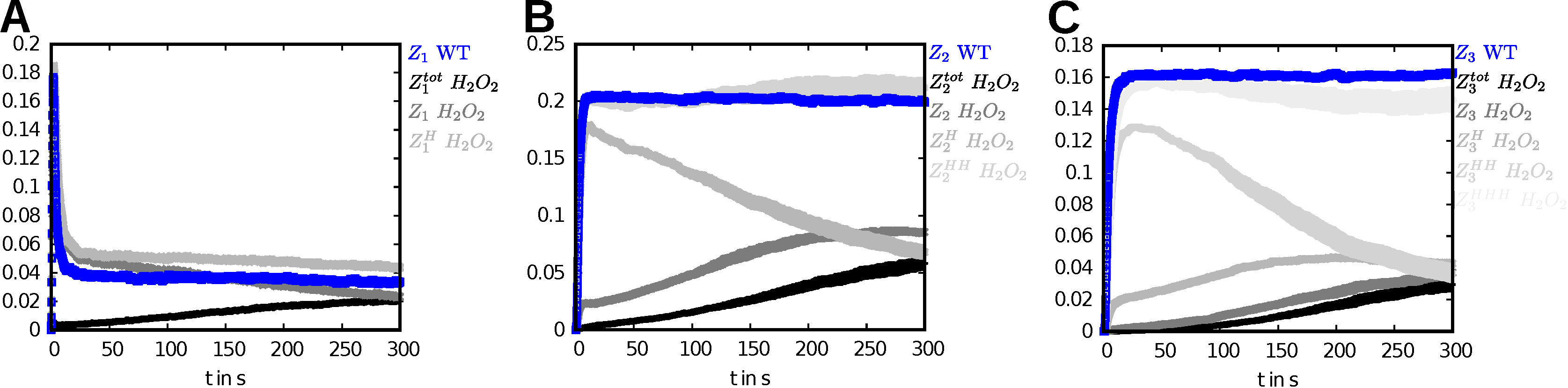}
\caption{\label{fig:ROSdefault}\textbf{Temporal analysis within the $\hzwei$ scenario} A: Temporal courses of the number of CRAC channel subunits of the WT scenario (blue) and of the $\hzwei$ scenario (black and gray) for $t \in [0 \, s, 300 \, s]$, B: temporal courses of tetramer CRAC channel states of the WT scenario (blue) and of the $\hzwei$ scenario (black and gray) for $t \in [0 \, s, 300 \, s]$, C: temporal courses of hexamer CRAC channel states of the WT scenario (blue) and of the $\hzwei$ scenario (black and gray) for $t \in [0 \, s, 300 \, s]$}
\end{figure}
First, we exemplarily analyzed the temporal course of all occurring $Z^j_i$ at default values. Figure \ref{fig:ROSdefault}A shows the total number of single CRAC channel subunits in the $\hzwei$ scenario (black) compared to the base case scenario (blue). In addition, the two possible configurations of CRAC channel subunits in the $\hzwei$ scenario are shown in gray. We see that within the temporal course, the total numbers of subunit states stay constant in both scenarios except for statistical fluctuations. In contrast, the composition of $Z^{tot}_1$ in the $\hzwei$ scenario changes within time: The number of WT subunits ($Z^0_1$) increases, while the number of ROS-inhibited subunits ($Z^{H}_1$) decreases.

The analysis of the tetramer and the hexamer CRAC channel states shows similar features (figure \ref{fig:ROSdefault}B and C): The total numbers of the respective channel configurations stays constant in both scenarios, while the composition of $Z^{tot}_i$ of the $\hzwei$ scenario changes: For small t a lot of total ROS-inhibited channels can be found. For larger t the number of total ROS-inhibited channels decreases, while the numbers of mixed channel states and of homomeric WT channel states increase.

Comparing the total numbers of $Z_i$ of the two different scenarios we see the following: There are less single CRAC channel subunits, roughly equal tetramer CRAC channels and more hexamer CRAC channel states in the base case scenario than in the $\hzwei$ scenario. These differences as well as the dynamic changes in the composition of $Z^{tot}_i$ occur due to the differences in the reaction rates between WT channels and partially or completely ROS-inhibited channels. While $\tilde{k_1}$ and $\tilde{k_2}$ are larger than $k_1$ and $k_2$, the rates for reactions between higher CRAC channel states $\tilde{k_3}$ to $\tilde{k_7}$ are smaller than for simple WT CRAC channel states $k_3$ to $k_7$ (cf. eq. (\ref{eq:mod_rates}) and table \ref{table:TabReactionRate}).

\begin{figure}
\includegraphics[scale=0.45]{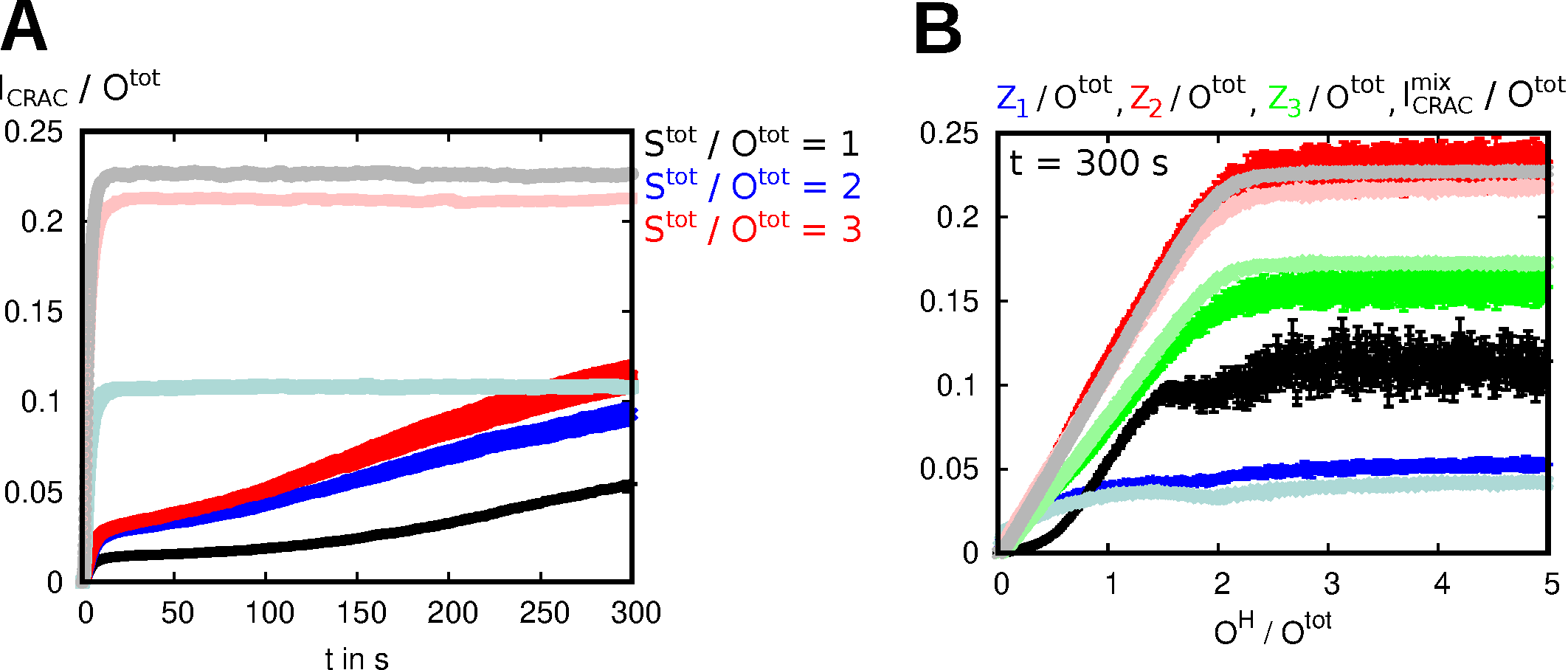}
\caption{\label{fig:ROSso}\textbf{$\hzwei$ scenario with different ratios $\stot / \otot$} \textbf{A:} Temporal course of $\icrach$ (cf. eq(\ref{eq:icracmix})) for different ratios of $\stot / \otot$ (base case scenario: shaded lines, $\hzwei$ scenario: bold lines) with fixed $\oh/\otot = 0.95$ and $\delta = 0.1$, \textbf{B:} values of the different CRAC channel states $Z_i$ and of $\icrach$ at $t = 300 \, s$ depending on the ratio of $\stot / \otot$ (base case scenario: solid lines, $\hzwei$ scenario: dashed lines)}
\end{figure}

\begin{figure}
\includegraphics[scale=0.45]{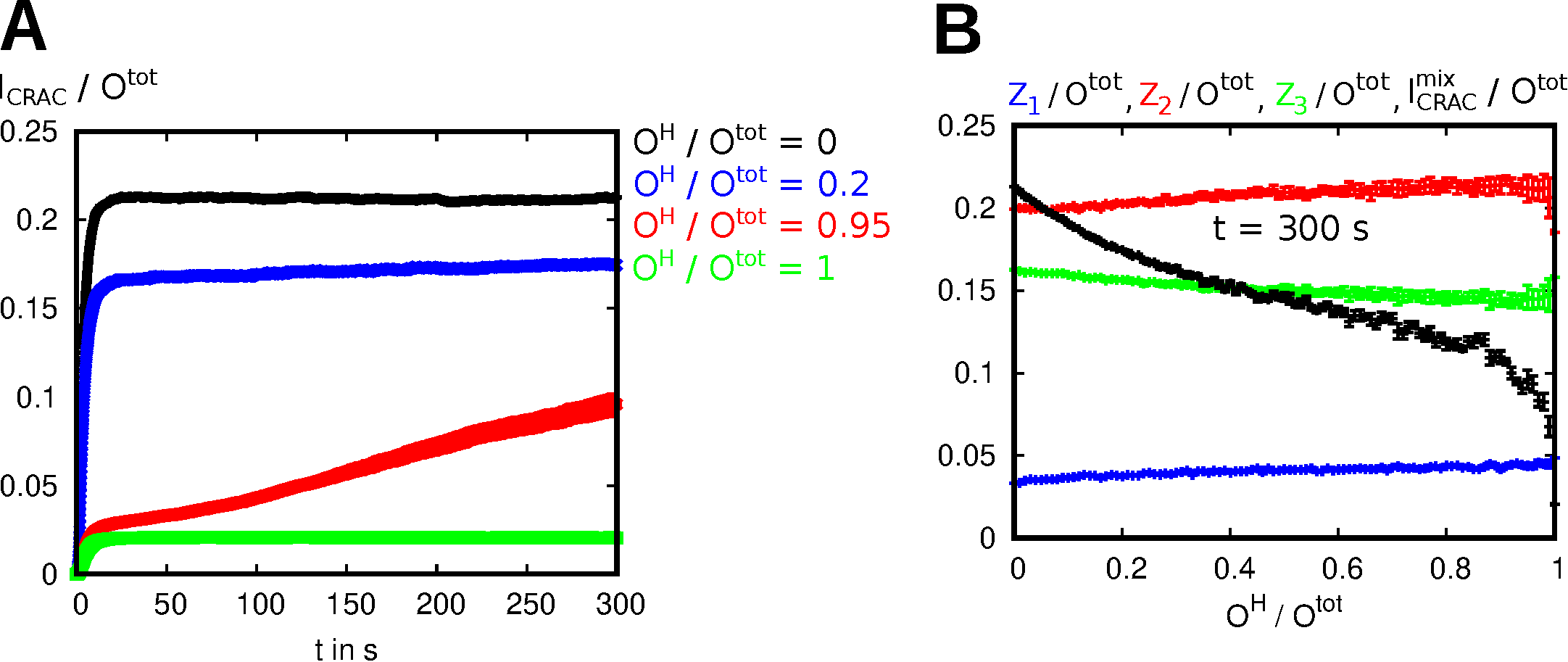}
\caption{\label{fig:ROSoho}\textbf{$\hzwei$ scenario with different ratios $\oh / \otot$} \textbf{A:} temporal course of $\icrach$ (cf. eq(\ref{eq:icracmix})) for different ratios of $\oh/\otot$ with fixed $\stot / \otot = 2$ and $\delta = 0.1$, \textbf{B:} values of the different CRAC channel states $Z_i$ and of $\icrach$ at $t = 300 \, s$ depending on the ratio $\oh/\otot$}
\end{figure}

\begin{figure}
\includegraphics[scale=0.45]{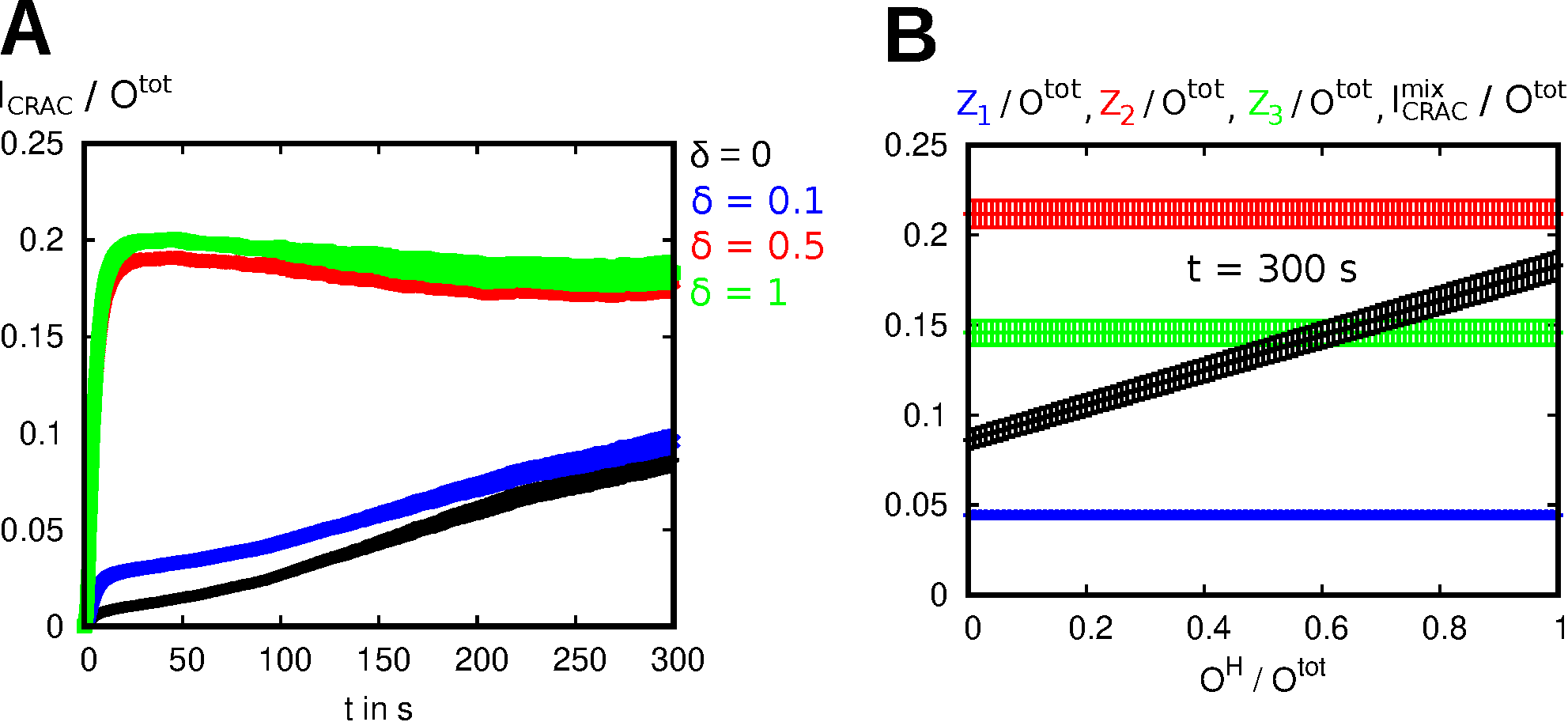}
\caption{\label{fig:ROSdelta}\textbf{$\hzwei$ scenario with different values of $\delta$} \textbf{A:} temporal course of $\icrach$ (cf. eq(\ref{eq:icracmix})) for different values of $\delta$ with fixed $S^{tot}/O^{tot} = 2$ and $\oh/\otot = 0.95$, \textbf{B:} values of the different CRAC channels $Z_i$ and of $\icrach$ at $t = 300 \, s$ depending on $\delta$}
\end{figure}

In the next step we varied these three parameters and analyzed the respective temporal courses of $\icrac$ (figs \ref{fig:ROSso}A, \ref{fig:ROSoho}A and \ref{fig:ROSdelta}A) and the steady state values of the different CRAC channel states $Z_i$ as well as of $\icrac$ normalized to $\otot$ (figs \ref{fig:ROSso}B, \ref{fig:ROSoho}B and \ref{fig:ROSdelta}B).

First, we analyzed the ratio $\stot / \otot$ while the other two parameters were fixed at their respective default values (figure \ref{fig:ROSso}B). As in the base case scenario, the highest $\cazweiplus$ influx is reached for $\stot / \otot \gtrsim 2.5$. Due to altered reaction parameters in the $\hzwei$ scenario, less hexameric channels and more tetrameric and dimeric channels are formed. The absolute value of $\icrac$ is significantly smaller ($\approx 35 \%$) in the $\hzwei$ scenario since the $\cazweiplus$ conductance of the ROS-inhibited ORAI1 channels (or channel parts) is just ten percent of the conductance of WT ORAI1 channels.

We compare the temporal course of $\icrac$ between the two scenarios for different ratios of $\stot / \otot$ (figure \ref{fig:ROSso}A, $\stot / \otot = 1$: black, $\stot / \otot = 2$: blue, $\stot / \otot = 3$: red). We see that in the base case scenario (shaded lines) $\icrac$ reaches a stable state after the first five seconds and in the $\hzwei$ scenario (bold lines) an intermediate plateau state is reached within the same time. While in the base case scenario this is the final steady state value, in the $\hzwei$ scenario $\icrach$ rises again after about 100 seconds. The first rise is due to diffusion of the proteins to the PMJ and the first CRAC channel formation. The second rise is due to the mixture of original and modified on- and off-rates and the high amount of ROS-inhibited ORAI1.

Analyzing the temporal course of $\icrach$ at different ratios of $\oh/\otot$ (cf. figure \ref{fig:ROSoho}A) at the fixed ratio $\stot / \otot = 2$ and for $\delta = 0.1$ shows, that the chosen value of $\oh/\otot$ affects whether the value of $\icrach$ at $t = 300\, s$ is the steady state value or not. Values of the case $\oh/\otot = 0$ (black) correspond to the base case scenario. $\oh/\otot = 1$ (green) correspond to all ORAI1 dimers interacted with ROS before building CRAC channels. $\oh/\otot = 0.2$ (blue) connotes that 20 $\%$ of ORAI1 dimers interacted with ROS before building CRAC channels. In all three cases the value of $\icrac$ approached a stable after about five to ten seconds. In contrast at a ratio of $\oh/\otot = 0.95$, which means that 95 $\%$ of the available ORAI1 interacted with ROS, $\icrach$ rises continuously within the 300 seconds. This is due to the fact that a large amount of ORAI1 features the modified on- and off-rates of the $\hzwei$ scenario.

Figure \ref{fig:ROSoho}B shows the values of $\icrach$ at $t = 300\, s$ and of $Z_i$ normalized to $\otot$ in dependence on the ratio $\oh/\otot$ ($\stot / \otot =2$, $\delta = 0.1$). The number of $Z_1$ increases, whereas the number of $Z_3$ decreases for higher $\oh / \otot$. The number of $Z_2$ stays at similar level within the whole interval.
This behavior combined with a ten percent contribution of $Z_i$ to $\icrach$ ($\delta = 0.1$) yields less $\icrach$ the more ORAI1 dimers interacted with ROS (i. e. the larger the ratio $\oh/\otot$).
In total the final value of $\icrach$ for $\oh/\otot = 1$ is equal to ten percent of the value of $\icrach$ for $\oh/\otot = 0$, which is as expected: With $\oh/\otot = 1$ equation (\ref{eq:icracmix}) simplifies to 
\begin{align}
\icrach(\oh/\otot = 1) = \delta \cdot \left(c_1 \cdot N_{Z_1} + c_2 \cdot N_{Z_2} + c_3 \cdot N_{Z_3}\right)\, ,
\end{align}
which is equal to multiplying equation (\ref{eq:icracBase}) with $\delta$.

Finally the influence of $\delta$ for fixed ratios $S^{tot}/O^{tot} = 2$ and $\oh/\otot = 0.95$ is examined (figure \ref{fig:ROSdelta}B). In general the steady state value of $\icrach$ increases with increasing $\delta$, which is as expected, since the $\cazweiplus$ influx depends linear on $\delta$ (cf. equation \ref{eq:icracmix}). 

Comparing different values of $\delta$ the temporal courses and absolute values of $\icrach$ differ much from each other (figure \ref{fig:ROSdelta}A). Large $\delta$ leads to a faster increase of $\icrach(t)$, which is obvious from the data before (cf. figures \ref{fig:ROSdefault}B and C): In the first seconds mainly total ROS-inhibited CRAC channel states are formed. For large $\delta$ these states contribute strongly to $\icrach$.

For $\delta = 0$ (black), which means CRAC channels including ROS preincubated ORAI1 dimers can not conduct any $\cazweiplus$ into the cell, $\icrach$ rises within the total analyzed time interval ($t \in [0\, s , 300\, s ]$). A similar behavior is seen for small delta ($\delta = 0.1$, blue). In the case that channels including ROS preincubated ORAI1 contribute half as strong to $\icrach$ as the channels consisting of WT ORAI1 only ($\delta = 0.5$, red) the temporal course of $\icrach$ is almost stable after about 30 seconds. 
If the mixed and the WT channels contribute in equal manner to $\icrach$ ($\delta = 1$, green), the same rises to a maximum after 30 seconds and than drops continuously again.

These differences and especially these instabilities in the course of $\icrac$ arise, because the compositions of the total numbers of CRAC channel states $Z^{tot}_i$ are not stable either. Due to the different on- and off-rates as well as the different cooperativity parameters $\alpha$ and $\beta$ for WT ORAI1 and WT CRAC channel states and for ROS preincubated ORAI1 and mixed CRAC channel states, the process of binding and unbinding of CRAC channel states stays dynamic. Therefore, also the predicted $\cazweiplus$ current, which is estimated by the numbers of different CRAC channel states is dynamic.
\\
\\

Summing up, the introduction of the new species, the ORAI1 dimers which reacted with ROS before building CRAC channels, leads to changes of the predicted $\cazweiplus$ influx into the cell:
\begin{itemize}
\item  As CRAC channels including ROS preincubated ORAI1 dimers contribute less to the $\cazweiplus$ current than simple WT CRAC channels, the total $\icrach$ drops largely, when a lot of ROS preincubated ORAI1 dimers are taken into account. 
\item Introducing the new species $\oh$ leads to dynamic binding and unbinding of the CRAC channel states and therefore the temporal course of $\icrach$ differs from the course seen in the base case scenario. \item Depending on the chosen set of parameters ($\stot / \otot$, $\oh/\otot$ and $\delta$), $\icrach$ either continuously rises within the examined 300 seconds, reaches a stable steady state value after less then 30 seconds, or drops again after reaching a  maximum value. 
\end{itemize}
Simultaneously, one feature of the base case scenario remains unchanged in the $\hzwei$ scenario:
If the number of available STIM1 dimers is at least 2.5 times larger than the total number of ORAI1 dimers, the highest values of $\icrach$ at $t = 300\, s$ are reached.

\section{Discussion}\label{Discussion}
With the help of a reaction-diffusion model we analyzed the STIM1-ORAI1 stoichiometries during CRAC channel formation, the influence of dynamic PM junctions and additionally focused on the effects of ROS inhibition.

Within the base case scenario, which correlates to WT ORAI1 in experiments we find that the highest value of the amplitude of $\icrac$ is reached when the ratio of STIM1 / ORAI1 equals or is larger than 2 in the analytical and numerical analysis and equals or is larger than 2.5 in the stochastic analysis. The difference occurs, since in the stochastic case diffusion of the proteins is taken into account and the minimal number of STIM1 does not suffice the free ORAI1 dimers, since they are possibly not at the same spot. 

Analyzing the numbers of active channels and the distribution of those, it is found that the most frequently occupied channel state is the hexameric channel state, followed by tetrameric channels for the core reaction system. Taking into account a further reaction between free ORAI1 dimers and fully open CRAC channel states (what we call \glqq stealing mechanism\grqq), the most dominant channel configuration is tetrameric followed by hexameric states. In this case the total amount of $\icrac$ is lower than without stealing mechanism. The dimeric channel subunit plays a minor role in contributing to $\cazweiplus$ influx into the cell. 
This analysis indicates, that the most likely channel configuration and with that the amount of $\cazweiplus$ influx can be controlled by the interplay between single CRAC channel subunits and already established CRAC channels.

Analyzing changes in the reaction rate parameters (on- and off-rate as well as cooperativity parameters $\alpha$ and $\beta$) shows, that for small off-rates the tetramer CRAC channel state configuration is more often occupied than the hexamer configuration in both cases, without and with stealing mechanism. For larger off-rates the situation changes: Considering the core reactions only, the hexamer configuration is more likely than the tetramer configuration. Considering the core reaction system with stealing mechanism, the tetramer configuration is still more often occupied, but the gap between the tetramer and the hexamer configuration is much smaller.
The influence of the on-rate cooperativity parameter $\alpha$ on the different channel states is negligible in the examined parameter range.
Large off-rate cooperativity parameter $\beta$ leads to less hexamer CRAC channel states than small $\beta$ and simultaneously more tetramer CRAC channel states.

Analytic and numeric analysis of the steady state value of $\icrac$ in dependence on the on- and off-rates as well as on the ratio of STIM1 / ORAI1 show, that increasing both the unbinding cooperativity and the ratio STIM1 / ORAI1 results in a slight decrease of $I_{CRAC}^{stat}$  after the highest value (for STIM1 / ORAI1 = 2) is reached. This is in good agreement with the exerimental results of \cite{kilch2013}.

In a further step the influence of ROS-inhibited channels was analyzed. Therefore the whole reaction-diffusion system was extended by a new type of species (the ROS preincubated ORAI1) and the formation of channel complexes was altered accordingly. Analysis show that still for the ratio of STIM1 / ORAI1 being equal or larger than  2.5 the largest $\cazweiplus$ influx is observed. In this case the maximum value of $\icrac$, which  is only $35 \%$ of the maximum value of the base case scenario, is reached. The introduced inhibition parameter $\delta$ influences $\icrac (t)$ for given ratios STIM1 / ORAI1 and $\oh/\otot$.
For large amounts of $\oh$ with respect to the total amount of ORAI1, $\icrac$ drops drastically.

Alansary et al. \cite{alansary2016} analyzed the influence of ROS on the amplitude of $\cazweiplus$ influx experimentally. In accordance to these experiments our analysis shows, that the drop in $\icrac$ for preincubated ORAI1 is not only due to alterations in the diffusion constant of ORAI1 and in the ORAI1-ORAI1 and the ORAI1-STIM1 interaction, but that there is another mechanism needed to explain the strong decrease of $\icrac$. In \cite{alansary2016} it is found that the interaction of two transmembrane domains lock the CRAC channel when preincubated with ROS. 
The introduction of the new channels into the model to account for this locking shows, that with this modification the drastic decrease of $\icrac$ can be explained, while only changing the reaction and diffusion parameters according to the changes found in the experiment does not yield such a large decrease in $\icrac$.

Recent experiments, which analyze the ORAI1 stoichiometry of CRAC channels demonstrate that fully functional ORAI channels exist as hexamers, in accorcance with the crystal structure \cite{hou2012, yen2016, cai2016}. Other experiments indicate that CRAC channels can occur as a a mixture of hexameric, tetrameric and dimeric channel complexes or as hexamers with different conducting states \cite{li2016, dynes2016}. A very recent report provides evidence for the existence of different conductance states of hexameric channels due to the ability of STIM1 dimers to crosslink neighbouring ORAI1 hexamers to provide a more efficient activation for a given number of STIM1 molecules \cite{zhou2018}.
Until now, there has been no model taking these new findings into account. We therefore close this gap and present a reaction-diffusion model in which we include three types of channels and therefore three types of conducting states: A CRAC channel subunit with only one ORAI1 dimer bound to STIM1 proteins with low conductance, an intermediate tetrameric CRAC channel with medium conductance and a fully open (i. e. hexameric) CRAC channel state with full conductance. Our model would lead to similar results if we assume the different conductance states to be due to hexameric channels with different number of bound STIM1 molecules.

With the help of the introduced \glqq stealing mechanism\grqq\, we can account for the dynamic reversible binding reactions between free STIM1 and ORAI1 as seen in \cite{wu2014}. In contrast to these findings in our model the free proteins remain within the junctional region.

Previous models \cite{peglow2013, hoover2011, wu2014, melunis2016} all assume tetramers as highest ORAI1 complexes. All studies evolve the best ratio of single STIM1 monomers to single ORAI1 monomers to be roughly 2:1. Also in our model, which is adapted to tetrameric and hexameric CRAC channels, the ratio of STIM1 / ORAI1 must be at least two or larger to gain highest $\cazweiplus$ influx.

While the models of \cite{peglow2013} and \cite{melunis2016} assume a static area as PMJ region, we allow a dynamic grow of the PMJ regions as seen in the cluster analysis of TIRF experiments.

In contrast to the previous models we can distinguish between the reaction-diffusion analysis of WT ORAI1 with STIM1 and the reaction-diffusion analysis of ROS-inhibited or mutated ORAI1 with STIM1. We can tune the quantity of inhibited proteins and therefore can predict the magnitude of $\cazweiplus$ influx in dependence on the ratio of preincubated ORAI1 to the total amount of ORAI1.

In summary the described model opens the possibility to predict $\cazweiplus$ influx into cells while modifying reaction and diffusion parameters, the amount of ROS-inhibited or mutated ORAI1 proteins and the predefinition of PMJ regions. So far, assessing and tuning these parameters experimentally is hard. Therefore our model can reveal new insights in the complex and dynamic stoichiometry of STIM1 and ORAI1 proteins during CRAC channel formation.

Addressing the Orai-STIM interaction during CRAC channel formation may also rise the question of how different STIM and Orai homologs (i. e. STIM2, ORAI2 and ORAI3) may influence the stoichiometry of the CRAC channels and the $\cazweiplus$ influx. For example recent studies \cite{saul2016} find that the CRAC channels consisting of a mixture of ORAI1 and Orai3 show redox insensitivity.
Further extensions of a model according to these homologs may offer new insights into the complex analysis of CRAC channel formation.

\subsection*{Author Contributions}
BS performed calculations and experiments, and analyzed the data. DA performed experiments. HR supervised and designed the theoretical aspects of the research, BAN advised on biological aspects and supervised the experiments. BS, DA, BAN, IB and HR wrote the manuscript.
\begin{acknowledgments}
This work was funded by the German Research Foundation (DFG) within the Collaborative Research Center SFB 1027, projects A3 to HR and C4 to BN and IB.
\end{acknowledgments}





\begin{thebibliography}{10}
\bibitem{hoover2011} Hoover P J and Lewis R S,
Stoichiometric requirements for trapping and gating of $\cazweiplus$ release-activated $\cazweiplus$ (CRAC) channels by stromal interaction molecule 1 (STIM1), Proc. Natl. Acad. Sci. U.S.A., 108, 32, 13299-13304 (2011)

\bibitem{berridge2000} Berridge M J, Lipp P and Bootman M D, The versatility and universality of calcium signalling, Nature Reviews Molecular Cell Biology, 1,  11-21 (2000)

\bibitem{clapham2007} Clapham, David E., Calcium Signaling, Cell, 131, 6, 1047-1058 (2007)

\bibitem{carafoli2002} Carafoli E, Calcium signaling: A tale for all seasons, PNAS, 99, 3, 1114-1122 (2002) 


\bibitem{falke2004} Falcke M, Reading the patterns in living cells — the physics of ca2+ signaling, Advances in Physics, 53, 3, 255-440 (2004)

\bibitem{lewis2001} Lewis R S, Calcium Signaling Mechanisms in T Lymphocytes, Annu. Rev. Immunol. 19, 497-521 (2001)

\bibitem{parekh2005} Parekh A B and Putney Jr. J W, Store-Operated Calcium Channels, 
Physiological Reviews, 85, 2, 757-810 (2005)

\bibitem{hoth1992} Hoth M and Penner R, Depletion of intracellular calcium stores activates a calcium current in mast cells, Nature, 355, 6358, 353-356, (1992)

\bibitem{muik2008} Muik M, Frischauf I, Derler I, Fahrner M, Bergsman J, Eder P, Schindl R, Hesch C, Polzinger B, Fritsch R, Kahr H, Madl J, Gruber H, Groschner K and Romanin C, Dynamic Coupling of the Putative Coiled-coil Domain of ORAI1 with STIM1 Mediates ORAI1 Channel Activation, The Journal of Biological Chemestry, 283, 12, 8014-8022 (2008)

\bibitem{luik2008} Luik R M, Wang B, Prakriya M, Wu M M and Lewis R S, Oligomerization of STIM1 couples ER calcium depletion to CRAC channel activation, Nature, 454(7203), 538-542 (2008)


\bibitem{derler2016} Derler I, Jardin I and Romanin C, Molecular mechanisms of STIM/Orai communication, Am J Physiol Cell Physiol, 310, C643-C662 (2016)

\bibitem{wu2011} Wu M M, Buchanan JA, Luik R and Lewis R S, $\cazweiplus$ store depletion causes STIM1 to accumulate in ER regions closely associated with the plasma membrane, The Journal of Cell Biology, 174, 6, 803-813 (2006)

\bibitem{liou2007} Liou J, Fivaz M, Inoue T and Meyer T, Live-cell imaging reveals sequential oligomerization and local plasma membrane targeting of stromal interaction molecule 1 after $\cazweiplus$ store depletion, PNAS, 104, 22, 9301-9306 (2007)

\bibitem{park2009} Park C Y, Hoover P J, Mullins F M, Bachhawat P, Covington E D, Raunser S, Walz T, Garcia K C, Dolmetsch R E and Lewis R S, STIM1 Clusters and Activates CRAC Channels via Direct Binding of a Cytosolic Domain to Orai1, Cell, 136, 876-890 (2009)

\bibitem{yuan2009} Yuan J P, Zeng W, Dorwart M R, Choi Y-J, Worley P F and Muallem S, SOAR and the polybasic STIM1 domains gate and regulate Orai channels, Nature Cell Biology, 11, 3, 337-343 (2009)

\bibitem{yen2016} Yen M, Lokteva L A and Lewis R S, Functional Analysis of Orai1 concatemers Supports a Hexameric Stoichiometry for the CRAC Channel, Biophysical Journal, 111, 1897-1907 (2016)

\bibitem{li2016} Li P, Miao Y, Dani A and Vig M, $\alpha$-SNAP regulates dynamic, on-site assembly and calcium selectivity of Orai1 channels, Molecular Biology of Cell, 27(16):2542-2553 (2016)

\bibitem{cai2016}
Cai X, Zhou Y, Nwokonko R M, Loktionova N A, Wnag X, Xin P, Trebak M, Wang Y and Gill D L,  The Orai1 Store-operated Calcium Channel Functions as a Hexamer, The Journal of Biological Chemistry, 291(50):25764-25775 (2016)

\bibitem{scrimgeour2009} Scrimgeour N, Litjens T, Ma L, Barritt G J and Rychkov G Y, Properties of Orai1 mediated store-operated current depend on the expression levels of STIM1 and Orai1 proteins, The Journal of Physiology, 587, 12, 2903-0918 (2009)

\bibitem{bogeski2010} Bogeski I, Kummerow C, Al-Ansary D, Schwarz E, Koehler R, Kozai D, Takahashi N, Peinelt C, Griesemer D, Bozem M, Mori Y, Hoth M and Niemeyer B A, Differential Redox Regulation of ORAI Ion Channels: A Mechanism to Tune Cellular Calcium Signaling,
Science Signaling, 3, 115 (2010) 

\bibitem{peglow2013} Peglow M, Niemeyer B A, Hoth M and Rieger H, Interplay of channels, pumps and organelle location in calcium microdomain formation, New Journal of Physics, 15, 27pp (2013)

\bibitem{li2011} Li Z, Liu L, Deng Y, Ji W, Du W, Xu P, Chen L and Xu T, Graded activation of CRAC channel by binding of different numbers of STIM1 to Orai1 subunits, Cell Research, 21, 305-315 (2011)

\bibitem{dynes2016} Dynes J L, Amacheslavsky A and Cahalan M D, Genetically targeted single-channel optical recording reveals multiple Orai1 gating states and oscillations in calcium influx, PNAS, 113 2, 440-445, (2016)

\bibitem{wu2014} Wu M M, Covington E D and Lewis R S,
Single-molecule analysis of diffusion and trapping of STIM1 and ORAI1 at endoplasmatic reticulum-plasma membrane junctions, Molecular Biology of the Cell, 25(22):3672-85 (2014)

\bibitem{shen2011} Shen W-W, Frieden Ma and Demaurex N, Remodeling of the endoplasmic retriculum during store-operated calcium entry, Biology of the Cell, 103 (8), 365-380 (2011)

\bibitem{sauc2015} Saüc S, Bulla M, Nunes P, Orci L, Marchetti A, Antigny F, Bernheim L, Cosson P, Frieden M and Demaurex N, STIML traps and gates Orai channesl without remodeling the cortical ER, Journal of Cell Science, 128, 1568-1579 (2015)

\bibitem{wu2006} Wu M M, Buchanan JA, Luik R M, and Lewis R S, $\cazweiplus$ store depletion causes STIM1 to accumulate in ER regions closely associated with the plasma membrane, The Journal of Cell Biology, 174, 6, 803-813 (2006)

\bibitem{malli2008} Malli R, Naghdi S, Romanin C and Graier W F, Cytosolic $\cazweiplus$ prevents the subplasmalemmal clustering of STIM1: an intrinsic mechanism to avoid $\cazweiplus$ overload, Journal of Cell Science, 121(019), 3133-3139 (2008)

\bibitem{melunis2016} Melunis J, Hershberg U, A spatially heterogenous Gillespie algorithm modeling framework that enables individual molecule history and tracking, Engineering Applications of Artificial Intelligence, 62, 304 - 311 (2017)

\bibitem{kilch2013}Kilch T, Alansary D, Peglow M,Dörr K, Rychkov G, Rieger H, Peinelt C and Niemeyer B A,
Mutations of the $\cazweiplus$-sensing Stromal Interaction Molecule STIM1 regulate $\cazweiplus$ influx by altered oligomerization of STIM1 and by destabilization of the $\cazweiplus$ channel Orai1
J. Biol. Chem. 288, 1653 (2013)

\bibitem{alansary2016} Alansary D, Schmidt B, Dörr K, Bogeski I, Rieger H, Kless A and Niemeyer B A, Thiol dependent intramolecular locking of Orai1 channels, Scientific Reports, 6, 33347 (2016)

\bibitem{zhou2015} Zhou Y, Wang X, Wang X, Loktionova N A, Cai X, Nwokonko R M, Vrana E, Wang Y, Rothberg B S and Gill D L, STIM1 dimers undergo unimolecular coupling to activate Orai1 channels, Nature Communications, 6:8395 (2015)

\bibitem{gillespie1976} Gillespie D T, A General Method for Numerically Simulating the Stochastic Time Evolution of Coupled Chemical Reactions,  Journal of Computational Physics 22, 403-434 (1976)

\bibitem{gibson1999} Gibson M A and Bruck J, Efficient Exact Stochastic Simulation of Chemical Systems with Many Species and Many
Channels, J. Phys. Chem. A, 104, 1876-1889 (1999)

\bibitem{hou2012} Hou X, Pedi L, Diver M M, and Long S B,Crystal structure of the calcium release-activated calcium channel Orai, Science, 338, 1308-1313 (2012)

\bibitem{zhou2018} Zhou Y, Nwokonko R M, Cai X, Loktionova N A, Abdulqadir R, Xin P, Niemeyer B A, Wang Y, Trebak M and Gill D L, Cross-linking of Orai1 channels by STIM proteins, PNAS, 115 15, E3398--E3407 (2018)

\bibitem{saul2016} Saul S, Gibhardt C S, Schmidt B, Lis A, Pasieka B, Conrad D, Jung P, Gaupp R, Wonnenberg B, Diler E, Stanisz H, Vogt T, Schwarz E C, Bischoff M, Herrmann M, Tschernig T, Kappl R, Rieger H, Niemeyer B A and Bogeski I, A calcium-redox feedback loop controls human
monocyte immune responses: The role of ORAI $\cazweiplus$ channels, Science Signaling, 9, 418 (2016)

\end{thebibliography}
\end{document}